\documentclass[11pt,a4paper]{article}
\usepackage{jcappub}
\usepackage{graphicx}
\newcommand{\simgt}{\lower.5ex\hbox{$\; \buildrel > \over \sim \;$}}
\newcommand{\simlt}{\lower.5ex\hbox{$\; \buildrel < \over \sim \;$}}

\title{Large Scale Structures in the Kinetic Gravity Braiding Model
That Can Be Unbraided}

\author{Rampei Kimura,}
\author{Kazuhiro Yamamoto}
\emailAdd{rampei@theo.phys.sci.hiroshima-u.ac.jp}
\emailAdd{kazuhiro@hiroshima-u.ac.jp}

\affiliation{
Department of Physical Science, Hiroshima University,
Higashi-Hiroshima 739-8526,~Japan}
\abstract{
We study cosmological consequences of a kinetic gravity braiding model, 
which is proposed as an alternative to the dark energy model.
The kinetic braiding model we study is characterized by a parameter 
$n$, which corresponds to the original galileon cosmological model 
for $n=1$. 
We find that the background expansion of the universe of the kinetic 
braiding model is the same as the Dvali-Turner's model, which reduces 
to that of the standard cold dark matter model with a cosmological
constant ($\Lambda$CDM model) for $n$ equal to infinity. 
We also find that the evolution of the linear cosmological perturbation
in the kinetic braiding model reduces to that of the $\Lambda$CDM 
model for $n=\infty$.
Then, we focus our study on the growth history of the linear density 
perturbation as well as the spherical collapse in the nonlinear 
regime of the density perturbations, which might be important in 
order to distinguish between the kinetic braiding model and the 
$\Lambda$CDM model when $n$ is finite.  
The theoretical prediction for the large scale structure is 
confronted with the multipole power spectrum of the luminous 
red galaxy sample of the Sloan Digital Sky survey. 
We also discuss future prospects of constraining the kinetic 
braiding model using a future redshift survey like the WFMOS/SuMIRe 
PFS survey as well as the cluster 
redshift distribution in the South Pole Telescope survey. 
}
\begin{document}
\maketitle
\def\bfx{{\bf x}}
%%%%%%%%%%%%%%%%%%%%%%%%%%
\section{Introduction}
%%%%%%%%%%%%%%%%%%%%%%%%%%
Cosmological observations of type Ia supernovae \cite{Riess,Perlmutter}, 
the cosmic microwave background anisotropies \cite{Spergel,Komatsu}, 
the large scale structure of galaxies \cite{Percival,sloan}, and 
clusters of galaxies \cite{Allen,Rapetti} symbolize the precision 
cosmology of the present-day. 
The universe is undergoing a phase of an accelerated expansion, which 
raises the most challenging problem in the era of the precision 
cosmology, called the dark energy problem. The cosmic accelerated 
expansion is ascribed to the presence of an extra component of 
universe with a negative pressure \cite{PeeblesRatra,Padmanabhan}.
The cosmological constant fits the cosmological observational data, 
however, the smallness of the value cannot be explained naturally 
\cite{Weinberg,Weinberg2}.
The quantum field theory predicts the existence of the vacuum energy, 
which corresponds to the cosmological constant $\Lambda$. However, 
it has been known that there is the fine tuning problem i.e., the fact that 
the observed value of the vacuum energy or the cosmological constant,
$\rho_{\Lambda}\simeq10^{-47}\mathrm{GeV}^4$, is much smaller orders 
of magnitude than the value that we expect in the quantum field theory.

An alternative to explain the accelerated expansion of the present universe 
is to modify the theory of gravity at long distance, such as the scalar-tensor 
theory which is an extension of general relativity by adding non-minimally 
coupled scalar field to gravity \cite{Amendola,Uzan,Chiba,Bartolo,Perrotta},
$f(R)$ gravity altering the Einstein-Hilbert action by a general function 
of Ricci scalar $R$ \cite{Carroll,Nojiri,Capozziello,HuSawicki,Starobinsky,
Tsujikawafr,Nojirib}, and the
Dvali-Gabadadze-Porrati (DGP) model developed in the context 
of the brane world scenario \cite{DGP1,DGP2}. 
However, these modified gravity models are not necessarily successful 
\cite{Song,Maartens,KM,Schmidt,frr}.

Recently, as an alternative to general relativity, there has been 
proposed the galileon gravity models, which are constructed by introducing
the scalar field with the self-interaction whose Lagrangian is invariant 
in the Minkowski space-time under the Galilean symmetry 
$\partial_{\mu}\phi \to \partial_{\mu}\phi+b_{\mu}$, 
which keeps equation of motion at the second order differential equation
\cite{GC,SAUGC,ELCP,CEGH,GGC,GBDT,DPGMGT,CG,GGIR,MGALG,CCGF,OCG,CSTOG,
Deffayet,DDEF,GALMG,RFFGM,SSSDBI,DBIGR,BGT1,BGT2,BTJC,APS,KMD,MKJMT,CSG,
GA}.
The simplest term of the self-interaction of the galileon model 
is $(\nabla\phi)^2 \square\phi$, which appears in the 4-dimensional 
effect theory of the DGP model. 
The galileon models generalized to a curved spacetime do not
necessarily possess the Galilean symmetry, however, it is possible
to keep the equation of motion at the second order differential equation.
The galileon model in the framework of the Brans-Dicke theory has been 
studied in \cite{GC,SAUGC,ELCP,CEGH,GGC,GBDT,DPGMGT}. 
In addition to these works, the authors in 
\cite{CG,GGIR,MGALG,CCGF,OCG,CSTOG,FeliceTsujikawa} investigated the 
cosmology of the covariant galileon field. 
These models can lead the self-accelerated expansion of the late-time 
universe. This nature is applied to construct the inflation
model \cite{GIKobayashi,MK} (cf.~\cite{GI,GIKobayashiII}). 
The galileon field equation of motion is the second order differential 
equation with respect to the time, which prevents the theory from 
appearing a new degree of freedom, and the perturbation of the theory 
does not suffer from ghost or instability problem. 
A remarkable property of the galileon model is the 
Vainshtein mechanism \cite{Vainshtein}.
The self-interaction term $(\nabla\phi)^2 \square\phi$ induces 
the decoupling of the galileon field $\phi$ from gravity at 
small scale. This allows the galileon theory to recover 
general relativity around a high density region, 
which ensures the consistency with the solar system experiments.

The kinetic braiding model is proposed inspired from the
galileon model \cite{Deffayet}, which introduced the 
extended self-interaction term $G(\phi,X)\square \phi$ 
minimally coupled to gravity, where $G(\phi,X)$ is a 
function of $\phi$ and $X$, which we defined 
$X=-g^{\mu\nu}\nabla_\mu \phi\nabla_\nu \phi/2$. 
In the present paper, we consider the case $G(\phi,X)\propto X^n$, 
where $n$ is the parameter. 
In the original work in \cite{Deffayet}, the case of $n=1$ is considered. 
Adopting an attractor solution for the galileon field,
we find that the background expansion for general value of $n$
realizes the Dvali-Turner model \cite{DT2003,KoyamaDGPn}, which is
the phenomenologically modified Friedmann equation,   
In the Dvali-Turner model, the background expansion reduces to 
that of the $\mathrm{\Lambda CDM}$ model for $n$ equal to infinity. 
We also find that the linear cosmological perturbations of the 
kinetic braiding model reduces to that of the 
$\mathrm{\Lambda CDM}$ model for $n=\infty$.
Thus, the kinetic braiding model connects the original
galileon model and the $\mathrm{\Lambda CDM}$ model by
the parameter $n$ at least as for the background expansion 
and the cosmological linear perturbations. 
This interesting feature is one of the reasons why we consider 
this model.

In general, when a modified gravity model shares a similar expansion 
history to those of dark energy models, there is a crucial 
difficulty in discriminating these models only with measurements 
of the background evolution. 
Therefore, it is worth examining the growth history of the 
large scale structure, because the degeneracy of the
background evolution could be broken (e.g., \cite{Uzanb,YS2007}).
In the present paper, we examine the evolution of 
density perturbations in the kinetic braiding model 
in the linear regime as well as the nonlinear regime 
by considering the spherical collapse model, which 
might be important in order to distinguish between the 
kinetic braiding model with a finite $n$ and the dark 
energy model as well as the $\Lambda$CDM model. 

The paper is organized as follows: In section \ref{background}, 
we briefly review the kinetic braiding model, where the 
background evolution and the stability condition of
the galileon field perturbation are summarized. 
In section \ref{snc},  
constraints from type Ia supernovae observations 
and the WMAP cosmic microwave background (CMB) 
anisotropy experiment are presented.
In section \ref{LinearPerturbation}, we investigate the linear 
evolution of the cosmological density perturbations.  
We show that the linear cosmological perturbations of the 
kinetic braiding model reduces to that of the 
$\mathrm{\Lambda CDM}$ model for $n=\infty$. 
We find an analytic formula of the growth rate and the growth 
index using  the quasi-static and sub-horizon approximation for 
the model with $n\simlt10$.  
In section \ref{PS}, we give a current constraint using the multipole power 
spectrum from the Sloan Digital Sky Survey (SDSS) luminous red galaxy 
(LGR) samples. A prospect of  constraining 
the kinetic braiding model from a future redshift survey is also considered.
In section \ref{VainshteinMechanism}, 
we investigate the behavior of the galileon field on small scales 
and the non-linear evolution of the density perturbation with the
spherical collapse approximation. We also show that the kinetic braiding 
model with small $n$ can be distinguished from the $\Lambda \mathrm{CDM}$ 
model using the galaxy cluster number count in a future survey.
Throughout the paper, we use units in which the speed of light the Planck
constant are unity, $c=\hbar=1$, and $M_{\rm Pl}$ is the reduced Planck mass
related with Newton's gravitational constant by $M_{\rm Pl}=1/\sqrt{8\pi G}$. 
We adopt the Hubble constant $H_0=100 h \mathrm{km/s/Mpc}$ with $h=0.7$, 
and the matter density parameter $\Omega_0h^2=0.1344$, 
unless stated otherwise.
We follow the metric signature convention $(-,+,+,+)$.

%%%%%%%%%%%%%%%%%%%%%%%%%%
\section{Kinetic Braiding Model}
\label{background}
%%%%%%%%%%%%%%%%%%%%%%%%%%
We consider the action that the galileon field $\phi$ is minimally coupled 
to gravity,
\begin{eqnarray}
  S=\int d^4x \sqrt{-g}\left[\frac{M_{\mathrm{Pl}}^2}{2}R
  +K(\phi,X)-G(\phi,X)\square \phi+\mathcal{L}_{\mathrm{m}} \right],
\label{Lagrangian}
\end{eqnarray}
where $R$ is the Ricci scalar, $K(\phi,X)$ and $G(\phi,X)$ are 
arbitrary functions of the galileon field and the kinetic term $X$, 
which is defined by 
$X=-g^{\mu\nu}\nabla_\mu\phi\nabla_\nu\phi /2$, %=-(\nabla\phi)^2/2$, 
$\square\phi=g^{\mu\nu}\nabla_\mu\nabla_\nu\phi$, and $\cal{L}_{\mathrm{m}}$ 
is the matter Lagrangian. 
The term $G(\phi,X)\square \phi$ is a general extension of 
the self-interaction term $X\square\phi$, %$(\nabla\phi)^2 \square\phi$. 
which is inspired 
%$(\nabla\phi)^2 \square\phi$ 
from a decoupling limit in DGP model %and also appears in 
and the low-energy string effective action \cite{DGP3,DGP4}. 
We consider the extended model motivated by 
its phenomenological consequences, as is shown below. 
 
First, the evolution equation of the galileon field is the 
second order differential equation.
The variation of the action 
with respect to $g^{\mu\nu}$ yields the equation for the
gravity,
\begin{eqnarray}
  M_{\mathrm{Pl}}^2 G_{\mu\nu}=T_{\mu\nu}^{(\phi)}+T_{\mu\nu}^{(m)},
\end{eqnarray}
where $T_{\mu\nu}^{(m)}$  is the energy momentum tensor of the matter 
and we defined 
\begin{eqnarray}
&&T_{\mu\nu}^{(\phi)}=K_X\nabla_\mu\phi\nabla_\nu\phi+g_{\mu\nu}K
+g_{\mu\nu}\nabla_\alpha G\nabla^\alpha\phi
\nonumber\\
&&~~~~~~~~~
-\left(
\nabla_\nu G\nabla_\mu\phi+\nabla_\mu G\nabla_\nu\phi\right)
-G_X\square\phi\nabla_\mu\phi\nabla_\nu\phi.
\end{eqnarray}
The galileon field equation is, 
\begin{eqnarray}
&&K_\phi+\nabla^\alpha(K_X\nabla_\alpha\phi)+2G_{\phi\phi}X-2G_\phi\square\phi
\nonumber
\\
&&~~
+G_{X\phi}\left(2X\square\phi
+2\nabla^\mu\phi\nabla_\mu\nabla_\nu\phi\nabla^\nu\phi\right)
\nonumber
\\
&&~~
+G_X\left[-(\square\phi)^2+(\nabla_\mu\nabla_\nu\phi)^2+R_{\mu\nu}\nabla^\mu\phi\nabla^\nu\phi
\right]
\nonumber
\\
&&~~
-G_{XX}\left(\nabla_\nu X\nabla^\nu X-\nabla^\mu\phi\nabla_\mu\nabla_\nu\phi\nabla^\nu\phi\square\phi
\right)=0, 
\label{GFieldEq}
\end{eqnarray}
where we defined  $G_\phi=\partial G(\phi,X)/\partial \phi$, 
$G_X=\partial G(\phi,X)/\partial X$, and so on.
In the present paper, we consider the case 
$K(\phi,X)$ and $G(\phi,X)$ are the function only of $X$, as 
given by eqs.~(\ref{K}) and (\ref{G}). 
In particular, in this case, the presence of the shift 
symmetry of the galileon field, $\phi \to \phi + c$, 
allows the existence of a Noether current, which is given by 
\begin{eqnarray}
  J_{\mu}=(K_X-2G_{\phi}-G_X \square \phi)\nabla_{\mu}\phi
 +G_X\nabla_{\mu}\nabla_{\nu}\phi \nabla^{\nu}\phi.
\label{Jmu}
\end{eqnarray}
In this case, the galileon field equation (\ref{GFieldEq}) simply 
reduces to $\nabla_{\mu} J^{\mu}=0$.

%%%%%%%%%%%%%%%%%%%%%%%%%%
\subsection{Background Evolution}
%%%%%%%%%%%%%%%%%%%%%%%%%%
In this subsection, we summarize the background evolution of this model. 
For the spatially flat Friedmann-Robertson-Walker metric,
\begin{eqnarray}
  ds^{2}=-dt^{2}+a^{2}(t)\delta_{ij}dx^{i}dx^{j},
\end{eqnarray}
the gravity equations are given by
\begin{eqnarray}
&&3M_{\mathrm{Pl}}^2H^2=\rho_{\phi}+\rho_m+\rho_r,
\label{MFE}
\\
&&-M_{\mathrm{Pl}}^2\left(2\dot{H}+3H^2\right)=p_{\phi}+p_r,
\label{MFE2}
\end{eqnarray}
where $H(=\dot a/a)$ is the Hubble parameter, $\rho_m$ and 
$\rho_r$ are the energy density of matter and radiation, 
respectively, $p_r$ is the pressure of the radiation, and
the energy density and the pressure of the galileon field are defined by
\begin{eqnarray}
  \rho_{\phi}&=&-K+K_X\dot{\phi}^2-G_{\phi}\dot{\phi}^2+3G_XH\dot\phi^3,\\
  p_{\phi}&=&K-G_{\phi}\dot{\phi}^2-G_X\dot\phi^2\ddot\phi, 
\end{eqnarray}
respectively. The galileon field equation is given by
\begin{eqnarray}
  &&K_{\phi}-(K_X-2G_\phi)(\ddot\phi+3H\dot\phi)-K_{\phi X}\dot{\phi}^2-K_{XX}\ddot{\phi}\dot{\phi}^2+G_{\phi\phi}\dot\phi^2
  \nonumber
  \\
  &&~~~~
  +G_{X\phi}\dot\phi^2(\ddot\phi-3H\dot\phi)
  -3G_X(2H\dot\phi\ddot\phi+3H^2\dot\phi^2+\dot H\dot\phi^2)
  -3G_{XX}H\dot\phi^3\ddot\phi=0.
\end{eqnarray}
Note that the energy-momentum conservation for the matter and radiation
remains the same as the standard one,
\begin{eqnarray}
  \dot{\rho}_m+3H\rho_m=0,
\\
  \dot{\rho}_r+4H\rho_r=0.
\end{eqnarray}

In this paper, we consider 
the following case,
\begin{eqnarray}
&&K(X)=-X,
\label{K}
\\
&&G(X)=M_{\mathrm{Pl}} \left(\frac{r_c^2}{M_{\mathrm{Pl}}^2}X \right)^n,
\label{G}
\end{eqnarray}
where $n$ and $r_c$ are the model parameters,  and 
$r_c$ is called the cross-over scale in the DGP model, 
which has a unit of length \cite{KoyamaMaartens}. 
This model gives us phenomenologically interesting features, as 
we will show below. 
This model corresponds to the original galileon model in \cite{Deffayet}
if $n=1$. 

The Lagrangian is invariant under a constant shift 
symmetry $\phi \to \phi + c$, 
and the charge density of the Noether current (\ref{Jmu}) can be written as
\begin{eqnarray}
  J_0=\dot{\phi}\left(3\dot{\phi}G_X H-1\right).
\end{eqnarray}
The galileon field satisfies $\dot{J_0}+3HJ_0=0$ and its solution is given by 
$J_0 \propto 1/a^3$. 
Therefore, $J_0$ approaches zero as the universe expands. It has been 
shown that the simplest attractor solutions is located at $J_0=0$ 
and has two branches, which are $\dot{\phi}=0$ and 
\begin{eqnarray}
\dot{\phi}={1 \over 3G_X H}.
\label{Att}
\end{eqnarray}
The case that $\dot{\phi}=0$ has ghostly perturbation
as we will see in the next section and a 
self-accelerating solution does not exist. 
Therefore, we pick the other branch 
(\ref{Att})  throughout this paper.
Then, we have
\begin{eqnarray}
  &&G_{XX}=2(n-1)\frac{1}{3H{\dot \phi}^3},\\
  &&G_{XXX}=4(n-1)(n-2)\frac{1}{3H{\dot \phi}^5},
\end{eqnarray}
and 
\begin{eqnarray}
  &&{\ddot \phi}=-\frac{1}{2n-1}\frac{{\dot \phi}{\dot H}}{H}.
\end{eqnarray}
Using (\ref{Att}) and the Hubble parameter as the present epoch 
(Hubble constant) $H_0=H(a_0)$, 
the cross-over scale is given by
\begin{eqnarray}
  r_c=\left(\frac{2^{n-1}}{3n}\right)^{1/2n}\left[\frac{1}{6(1-\Omega_0-\Omega_{r0})}\right]^{(2n-1)/4n} H_0^{-1},
\label{crossover}
\end{eqnarray}
where $\Omega_0$ and $\Omega_{r0}$ are the density parameter of the matter and radiation,
respectively, at present.
In this case, we obtain the modified Friedmann equation,
\begin{eqnarray}
  \left(\frac{H}{H_0}\right)^2=(1-\Omega_{0}-\Omega_{r0})
  \left(\frac{H}{H_0}\right)^{-\frac{2}{2n-1}}+\Omega_{0}a^{-3}+\Omega_{r0}a^{-4}.
\label{DT}
\end{eqnarray}
We also find that the effective equation of state of the galileon field
can be written as 
\begin{eqnarray}
  w_{\mathrm{eff}}={p_\phi\over\rho_\phi}&=&\frac{2\dot{H}}{3(2n-1)H^2}-1 \\
  &=&-\frac{6n+\Omega_{r}(a)}{3(2n-\Omega_{m}(a)-\Omega_{r}(a))},
\end{eqnarray}
where $\Omega_{\phi}(a)=\rho_{\phi} / 3M_{\mathrm{Pl}}^2H^2$, 
$\Omega_{\rm m}(a)=\rho_{m} / 3M_{\mathrm{Pl}}^2H^2$, 
and $\Omega_{\rm r}(a)=\rho_{r} / 3M_{\mathrm{Pl}}^2H^2$,
which satisfy $\Omega_{\phi}(a)+\Omega_{\rm m}(a)+\Omega_{\rm r}(a)=1$. 
In the derivation of the expression for $w_{\mathrm{eff}}$, we used 
\begin{eqnarray}
  \frac{\dot{H}}{H^2}=-\frac{(2n-1)(3\Omega_{m}(a)+4\Omega_{r}(a))}{2(2n-\Omega_{m}(a)-\Omega_{r}(a))}.
\end{eqnarray}
The effective equation of state has the asymptotic values,
\begin{eqnarray}
    w_{\mathrm{eff}}\simeq \left\{ 
\begin{array}{cr}
-\displaystyle{\frac{6n+1}{3(2n-1)}} &{\rm (radiation~dominated~era)}, \\
-\displaystyle{\frac{6n}{3(2n-1)}} &{\rm (matter~dominated~era)},\\
-\displaystyle{1} &{\rm (scalar~field~dominated~era)}.\\
\end{array} \right.
\end{eqnarray}

The modified Friedmann equation (\ref{DT}) is known as the Dvali-Turner 
model, which was proposed as a phenomenologically modified
model of the DGP model \cite{DT2003,KoyamaDGPn}.
At early stage of the cosmological expansion after the matter-dominated
era, the matter energy density dominates the right-hand-side of 
the modified Friedmann equation (\ref{MFE}), which reduces to the
usual one, $3M_{\mathrm{Pl}}^2H^2 \simeq \rho_m$. 
At late time, the matter energy density can be neglected, 
therefore the first term dominates the right-hand-side of the 
Friedmann equation (\ref{MFE}). This term acts like the cosmological 
constant and asymptotically approaches the de Sitter expansion. 
This can be seen from that the effective equation of state approaches 
$-1$ for any $n$ in the future. 

There are two special cases $n=1$ and $n=\infty$ in background evolution. 
For $n=1$, we can solve the Friedmann equation for $H^2$,
\begin{eqnarray}
  &&\left(\frac{H}{H_0}\right)^2=\frac{1}{2}
\biggl[\Omega_{0}a^{-3}+\Omega_{r0}a^{-4}
%  \nonumber
%  \\
%  &&~~~~~~~~~~~~
+\sqrt{\left(\Omega_{0}a^{-3}
    +\Omega_{r0}a^{-4}\right)^2+4(1-\Omega_{0}-\Omega_{r0})}\biggr].
\end{eqnarray}
Note that this solution is the same as the tracker solution in \cite{OCG}.
Here the matter density parameter is given by 
\begin{eqnarray}
  \Omega_{\rm m}(a)=\frac{2 \Omega_0 a^{-3}}{\Omega_{0}a^{-3}+\Omega_{r0}a^{-4}+\sqrt{\left(\Omega_{0}a^{-3}
    +\Omega_{r0}a^{-4}\right)^2+4(1-\Omega_{0}-\Omega_{r0})}}.
\end{eqnarray}
During matter-dominated era, $\Omega_{\rm m}(a) \simeq 1$ gives $w_{\mathrm{eff}}=-2$. 
For large $n$, the Friedmann equation can be written as 
\begin{eqnarray}
  \left(\frac{H}{H_0}\right)^2 \simeq 1-\Omega_{0}-\Omega_{r0}
  +\Omega_{0}a^{-3}+\Omega_{r0}a^{-4}.
\end{eqnarray}
This is almost the same as the Friedmann equation of the 
$\Lambda\mathrm{CDM}$ model.   

\begin{figure}[t]
  \begin{tabular}{cc}
   \begin{minipage}{0.5\textwidth}
    \begin{center}
     \includegraphics[scale=1.4]{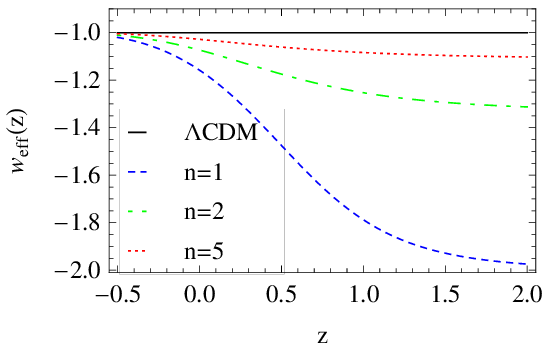}
    \end{center}
   \end{minipage}
   \begin{minipage}{0.5\textwidth}
    \begin{center}
     \includegraphics[scale=1.4]{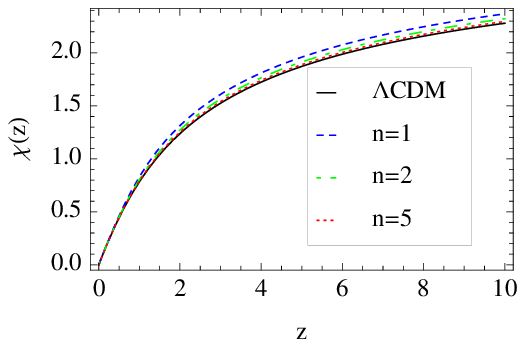}
    \end{center}
   \end{minipage}
  \end{tabular}
  \caption{Left panel: The effective equation of state $w_{\mathrm{eff}}$ 
  as a function of redshift for $\mathrm{\Lambda CDM}$ (solid curve) 
  and the kinetic braiding mode with $n=1$ (dashed curve), 
  $n=2$ (dash-dotted curve), and $n=5$ (dotted curve), respectively. 
  Right panel: The comoving distance $\chi(z)$, normalized by $H_0$, 
  as a function of redshift for $\mathrm{\Lambda CDM}$ and this model.
  }
  \label{fig1}
\end{figure}

The numerical calculation of the effective equation of state 
$w_{\mathrm{eff}}$ for 
different parameters $n$ is shown in figure~\ref{fig1} (left panel).  
As is expected, $w_{\mathrm{eff}}$ approaches $-1$ for large $n$. 
The right panel of figure~\ref{fig1} plots the comoving distance 
$\chi(z)=\int_0^z dz'/H(z')$. 
As $n$ becomes large, the background evolution approaches
that of the $\Lambda \mathrm{CDM}$ model.
  
%%%%%%%%%%%%%%%%%%%%%%%%%%
\subsection{Stability Condition}
\label{stability}
\begin{figure}[t]
  \begin{center}
    \includegraphics[scale=1.6]{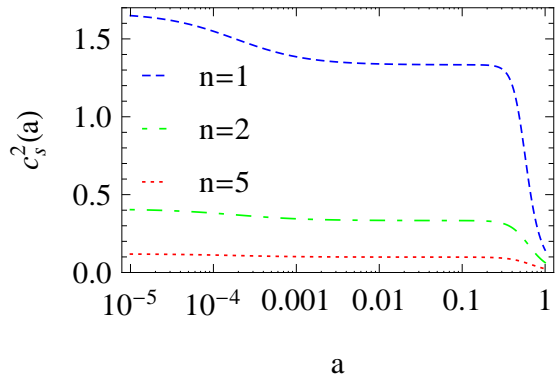}% 
    \caption{The sound speed of perturbations for $n=1$ (dashed curve), 
      $n=2$ (dash-dotted curve), and $n=5$ (dotted curve), 
      from the top to bottom, as a function of scale factor.
      \label{fig2} {}}
  \end{center}
\end{figure}

%%%%%%%%%%%%%%%%%%%%%%%%%%
Let us study the stability of the model by perturbatively 
expanding the action of the galileon field at second order 
as $\phi \to \phi + \delta\phi$. 
We neglect the metric perturbation, then the quadratic action becomes
\begin{eqnarray}
  \delta S^{(2)}=\frac{1}{2}\int d^4x \sqrt{-g}\left[\kappa(a) \dot{\delta\phi}^2 
- \frac{\beta(a)}{a^2} (\partial_i \delta\phi)^2 \right],
\label{stabilityAction}
\end{eqnarray}
where
\begin{eqnarray}
  \kappa(a)&=&K_X+K_{XX}\dot{\phi}^2-2G_\phi-G_{X\phi}\dot\phi^2
\nonumber
\\
&&~~~~~~~~+6G_XH\dot\phi+3G_{XX}H\dot\phi^3+12\pi GG_X^2\dot\phi^4,
\\
  \beta(a)&=&K_X-2G_\phi+G_{X\phi}\dot\phi^2+2G_X(\ddot\phi+2H\dot\phi)
\nonumber
\\
&&~~~~~~~~+G_{XX}\dot\phi^2\ddot\phi-4\pi GG_X^2\dot\phi^4.
\end{eqnarray}
To avoid ghost and instability, we require 
$\kappa(a)>0$ and $c_s^2=\beta(a) / \kappa(a) >0$, respectively. 
One of the attractor solution, $\dot{\phi}=0$, has obviously ghostly 
perturbation since $\kappa=-1 <0$ in our model. 
%In the case $\dot{\phi}=1 / 3G_X H$, 
Using the attractor condition (\ref{Att}),
$\kappa(a)$, $\beta(a)$, 
and $c_s^2(a)$ can be written in terms of 
$\Omega_{\rm m}(a)$  and $\Omega_{\rm r}(a)$ as 
\begin{eqnarray}
  \kappa(a)&=&(2n-\Omega_{\rm m}(a)-\Omega_{\rm r}(a)),
\\
  \beta(a)&=&\frac{n(5\Omega_{\rm m}(a)+6\Omega_{\rm r}(a))
      -(\Omega_{\rm m}(a)+\Omega_{\rm r}(a))^2}{3(2n-\Omega_{\rm m}(a)-\Omega_{\rm r}(a))} ,
\\
  c_s^2(a)&=&\frac{n(5\Omega_{\rm m}(a)+6\Omega_{\rm r}(a))-(\Omega_{\rm m}(a)+\Omega_{\rm r}(a))^2}
  {3(2n-\Omega_{\rm m}(a)-\Omega_{\rm r}(a))^2}.
\label{cs2}
\end{eqnarray}
Throughout the evolution of the universe, the matter density parameter 
$\Omega_{\rm m}(a)$ is always less than unity, therefore, we require $n>1/2$ 
to avoid ghost instability, $\kappa>0$. In figure~\ref{fig2}, 
the evolution of the sound speed of perturbations is plotted. 
The asymptotic behavior of the sound speed is given by
\begin{eqnarray}
  c_{s}^2\simeq \left\{ \begin{array}{lr}
\displaystyle{\frac{6n-1}{3(2n-1)^2}} & {\rm (radiation-dominated~era)}, \\
\displaystyle{\frac{5n-1}{3(2n-1)^2}} & {\rm (matter-dominated~era)}.\\
\end{array} \right.
\end{eqnarray}
%Therefore, the superluminarity can be avoided for $n>(9+\sqrt{33})/12 \simeq 1.23$.

It is worthy to note the fact that the sound speed of the 
galileon field perturbation becomes zero for $n=\infty$. 
In section 4, we show that the 
linear cosmological perturbation reduces to that of the
$\Lambda{\rm CDM}$ model.
In a case of large but finite value of $n$, the time-dependence
of the galileon field perturbation will become important 
even on sub-horizon scales.
For a case of small value of $n$, e.g., $n\simlt 10$, 
we can use the quasi-static approximation safely on 
small scales, which is demonstrated in section 4. 

%%%%%%%%%%%%%%%%%%%%%%%%%%
\section{Constraint from Supernovae and CMB distance Observation}
\label{snc}
\begin{figure}[]
  \begin{tabular}{cc}
   \begin{minipage}{0.5\textwidth}
    \begin{center}
     \includegraphics[scale=1.4]{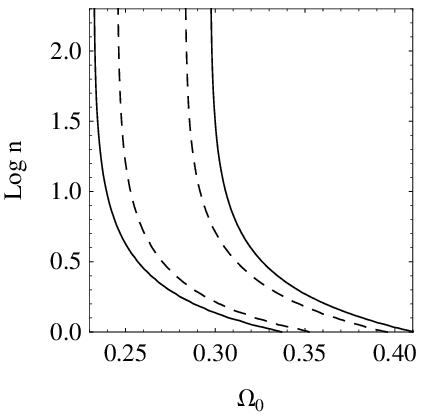}
    \end{center}
   \end{minipage}
   \begin{minipage}{0.5\textwidth}
    \begin{center}
     \includegraphics[scale=1.4]{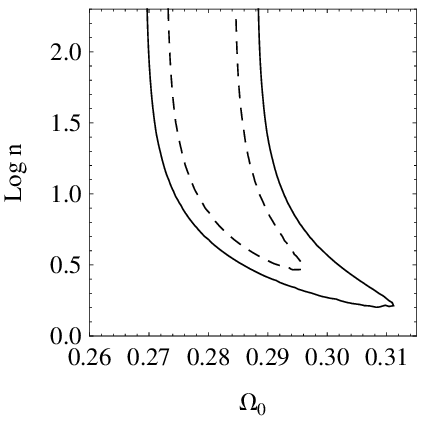}
    \end{center}
   \end{minipage}
  \end{tabular}
  \caption{The left panel is the contour of $\Delta\chi^2_{\mathrm{SN}}$ 
on the plane $\Omega_0$ and $n$ for the kinetic braiding model. 
The dashed curve and the solid curve are the $1$ $\sigma$ and $2$ 
$\sigma$ confidence contours, respectively.
The right panel is the same but of $\Delta\chi^2_{\mathrm{CMB}}$. 
  \label{figsn}}
\end{figure}

%%%%%%%%%%%%%%%%%%%%%%%%%%
The SCP Union2 Compilation is a collection of 557 type Ia supernovae data 
whose range of the redshift is $0.015 < z < 1.4$ \cite{SCP}. 
The distance modulus, which is the 
difference between apparent and absolute magnitude of the object, is given by
\begin{eqnarray}
  \mu=5\log \left(\frac{d_L(z)}{Mpc}\right)+25,
\end{eqnarray}
where $d_L(z)$ is the luminosity distance,
\begin{eqnarray}
  d_L(z)=(1+z)\int_0^zH^{-1}(z')dz'=(1+z)\chi(z).
\end{eqnarray}
The best-fit value of the present matter density $\Omega_0$ and $n$ 
can be determined 
by $\chi^2_{\mathrm{SN}}$ defined by
\begin{eqnarray}
  \chi^2_{SN}=\sum_{i=1}^{557}\left[\frac{(\mu(z_i)-\mu_{obs}(z_i))^2}{\sigma^2_{obs}(z_i)} 
\right],
\end{eqnarray}
where $z_i$,  $\mu_{obs}(z_i)$, and $\sigma_{obs}(z_i)$ are the redshift, 
the distance modulus, and the error  of the $i$-th observed Ia SN. 
%In our analysis, we adopt $h=0.7$. 
The left panel of figure \ref{figsn} shows the confidence contour of 
$\Delta\chi^2_{\mathrm{SN}}$ on 
the plane $\Omega_0$ and $n$ for the kinetic braiding model.
We find the minimum of the $\chi^2_{\mathrm{SN}}$ is respectively given by 
$\chi^2_{\mathrm{SN}}=542.7$ 
for $\Lambda$CDM model and $\chi^2_{\mathrm{SN}}=543.3$ for the kinetic
braiding model with $n=1$.
The kinetic braiding model with $n=1$ requires the higher 
matter density parameter.
A similar result is found in a recent paper \cite{OCG}.

The Wilkinson microwave background anisotropy (WMAP) observation
also provide a constraint that relies on the distance to the 
last scattering surface. We here adopt the method using 
the shift parameter $R$, the acoustic scale $l_A$, and the redshift 
of the decoupling epoch $z_*$ \cite{WangMukherjee}. 
In the spatially flat universe, 
the acoustic scale and the shift parameter are 
written $l_A=\pi \chi(z_*)/r_s(z_*)$ and 
$R=\sqrt{\Omega_0H_0^2}\chi(z_*)$, respectively, 
where $\chi(z)$ is the comoving distance and $r_s$ is
the comoving sound horizon at the decoupling epoch 
\cite{HuSugiyama}. Then, we define the chi squared 
(see \cite{Komatsu,OCG} for details)
\begin{eqnarray}
\chi^2_{\rm CMB}=\sum_{i,j}(x_i-d_i){\cal C}_{ij}^{-1}(x_j-d_j),
\end{eqnarray}
where $x_i=(l_A,R,z_*)$, $d_i=(302.09,~1.725,~1091.3)$, and
the inverse covariance matrix
\begin{eqnarray}
{\cal C}_{ij}^{-1}=\left(
\begin{array}{ccc}
~~2.305~~&~~29.698~~&~~-1.333~~
\\
~~29.698~~&~~6825.27~~&~~-113.18~~
\\
~~-1.333~~&~~-113.18~~&~~3.414~~
\end{array}
\right).
\end{eqnarray}
In this analysis, we adopted $\Omega_{r0}
=4.17\times 10^{-5} h^{-2}$.
The right panel of figure \ref{figsn} shows 
$1$ $\sigma$ and $2$ $\sigma$ confidence contours 
of $\Delta \chi^2_{\mathrm{CMB}}$.
Our result is consistent with \cite{OCG} for the case $n=1$.
The higher $n$ models better match the CMB distance observation.

%%%%%%%%%%%%%%%%%%%%%%%%%%
\section{Linear Cosmological Perturbations}
\label{LinearPerturbation}
%%%%%%%%%%%%%%%%%%%%%%%%%%
%\subsection{basic formulas}
In this section, we consider the linear evolution of cosmological 
perturbations. 
Hereafter, we consider the era after the matter domination.
The metric perturbation in the conformal Newtonian gauge is given by
\begin{eqnarray}
  ds^{2}=-(1+2\Psi)dt^{2}+a^{2}(t)(1+2\Phi)\delta_{ij}dx^{i}dx^{j}.
\end{eqnarray}
% The perturbation of the matter density and the galileon field is, 
% respectively, 
% defined by $\rho(x,t)=\rho(t)\left(1+\delta(x,t)\right)$ and 
% $\phi(x,t)=\phi(t)(1+\varphi(x,t))$. 
We write the perturbation of the galileon field by $\phi(x,t)
=\phi(t)+\delta \phi(x,t)$. Hereafter, $\dot\phi$ means $\dot\phi(t)$. 
In appendix A, we summarized the perturbation equations in the 
general case. In our model, ~(\ref{K}) and (\ref{G}), 
the following equations are obtained.

The $(0,0)$ component of the Einstein equation %(\ref{fullPertE00}) is 
yields
\begin{eqnarray}
&&2M_{\mathrm{Pl}}^2 \left[-3H(\dot\Phi-H\Psi)+{1\over a^2}\nabla^2\Phi\right]
\nonumber
\\
&&~~~~~~
=-K_X\delta X
-G_X\left(3\dot\phi^3\dot\Phi-12H\dot\phi^3\Psi+9H\dot\phi^2\dot{\delta\phi}
-{\dot\phi^2\over a^2}\nabla^2\delta\phi\right)
\nonumber
\\
&&~~~~~~
-3G_{XX}H\dot\phi^3\delta X-\delta\rho,
\label{fullPertE00} 
\end{eqnarray}
where we defined 
\begin{eqnarray}
\delta X=\dot\phi\dot{\delta\phi}-\dot\phi^2\Psi,
\label{defdeltaX}
\end{eqnarray}
and $\delta\rho$ is the matter density perturbation.
The $(i,j)$ component of the Einstein equation %(\ref{fullPertEij}) 
yields
\begin{eqnarray}
&&2M_{\mathrm{Pl}}^2 \left[(3H^2+2\dot{H})\Psi+H\dot{\Psi}-\ddot{\Phi}-3H\dot{\Phi} \right]
\nonumber
\\
&&~~
=K_{X}\delta X+G_{X}\left(\dot{\phi}^3\dot{\Psi}-\dot{\phi}^2\ddot{\delta\phi} 
+4\dot\phi^2\ddot{\phi}\Psi-2\dot\phi\ddot\phi\dot{\delta\phi}\right)
-G_{XX}\dot{\phi}^2\ddot{\phi}\delta X,
\label{fullPertEij} 
\end{eqnarray}
and 
\begin{eqnarray}
\Psi+\Phi=0.
\label{fullPertEtl}
\end{eqnarray}
The $(0,i)$ component yields
\begin{eqnarray}
&2M_{\mathrm{Pl}}^2 \left(\dot\Phi-H\Psi\right)
=-K_{X}\dot\phi\delta\phi-G_X \dot\phi^2\left(
\dot\phi\Psi-\dot{\delta\phi}+3H\delta\phi\right)+\delta q,
\label{fullPertE0i}
\end{eqnarray}
where $\delta q$ describes the velocity field (see appendix A). 
The perturbed equation for the galileon field %(\ref{fullPertField}) 
becomes
\begin{eqnarray}
~\hspace{-1cm}
&&-K_{X}\left[3\dot{\phi}\dot{\Phi}-\dot{\phi}\dot{\Psi}-2(\ddot{\phi}
+3H\dot{\phi})\Psi+\ddot{\delta\phi}+3H\dot{\delta\phi}
-\frac{1}{a^2}\nabla^2\delta\phi \right]
\nonumber
\\
&&
-3G_{XXX}H\dot\phi^3\ddot\phi\delta X
-G_{X}\biggl[3\dot{\phi}^2\ddot{\Phi}+6(\ddot{\phi}+3H\dot{\phi})\dot{\phi}\dot{\Phi}
-9H\dot{\phi}^2\dot{\Psi}
\nonumber
\\
&&
-12\biggl\{(\dot{H}+3H^2)\dot{\phi}^2
+2H\dot{\phi}\ddot{\phi} \biggr\}\Psi
-\frac{\dot{\phi}^2}{a^2}\nabla^2\Psi
+6H\dot{\phi}\ddot{\delta\phi}
\nonumber
\\
&&
+6\biggl\{H\ddot{\phi}+(\dot{H}+3H^2)\dot{\phi} \biggr\}\dot{\delta\phi}
-\frac{2}{a^2}(\ddot{\phi}+2H\dot{\phi})\nabla^2\delta\phi
 \biggr]
\nonumber
\\
&&
-G_{XX}\biggl[3\dot{\phi}^3\ddot{\phi}\dot{\Phi}-3H\dot{\phi}^4\dot{\Psi}
-3\biggl\{8H\dot{\phi}^3\ddot{\phi}+(\dot{H}+3H^2)\dot{\phi}^4\biggr\}\Psi
\nonumber
\\
&&
+3H\dot{\phi}^3
\ddot{\delta\phi}
+3\biggl\{5H\dot{\phi}^2\ddot{\phi}+(\dot{H}+3H^2)\dot{\phi}^3\biggr\}
\dot{\delta\phi}-\frac{\dot{\phi}^2\ddot{\phi}}{a^2}\nabla^2\delta\phi\biggr]
=0.
\label{fullPertField}
\end{eqnarray}
The perturbed equation for the matter component is described in appendix A.
Instead of solving the evolution equation of the matter perturbation, 
we use the following equation \cite{ELCP}, which is obtained by combining 
eqs.~(\ref{fullPertE00}) and (\ref{fullPertE0i}),
\begin{eqnarray}
\rho\Delta_c&=&-2M_{\mathrm{Pl}}^2 {\nabla^2\Phi\over a^2}
-K_{X}\left(\delta X+3H\dot\phi\delta\phi\right)
\nonumber
\\
&&
+G_{X}\left(-3\dot\phi^3\dot\Phi+9H\dot\phi^3\Psi-6H\dot\phi^2
\dot{\delta\phi}-9H^2\dot\phi^2\delta\phi+{\dot\phi^2}{\nabla^2\delta\phi\over a^2}\right)
\nonumber
\\
&&-3G_{XX}H\dot\phi^3\delta X,
\end{eqnarray}
where $\Delta_c$ is the gauge invariant matter density contrast.

%%%%%%%%%%%%%%%%%%%%%%%%%%%%%%
\subsection{Limit of $n=\infty$}
\label{secInf}
%%%%%%%%%%%%%%%%%%%%%%%%%%%%%%
In section 3, we showed that the background expansion of the 
kinetic braiding model approaches that of the $\Lambda$CDM model 
as $n$ becomes large. 
In this subsection, we show that the linear cosmological perturbation 
of the kinetic braiding model also reduces to that of the cosmological 
constant model for $n=\infty$.

{}From the attractor condition (\ref{Att}), in the limit of  $n=\infty$, 
we find that $\dot\phi$ is constant, 
\begin{eqnarray}
  &&{\dot \phi}=\left[6(1-\Omega_0)\right]^{1/2}H_0M_{{\rm Pl}}. 
\end{eqnarray}
In this case, the leading term 
expanded eq.~(\ref{fullPertField}) in terms of $1/n$ gives
\begin{eqnarray}
  &&\dot {\delta X}+3H\delta X=0,
\label{eqdelataX}
\end{eqnarray}
where we used (\ref{fullPertEtl}) and $\dot {\delta X}=\dot\phi
\ddot{\delta\phi}-\dot\phi^2\dot \Psi$. Then, the solution of 
(\ref{eqdelataX}) is $\delta X=C/a^3$, where $C$ is an integration
constant, which is determined by initial condition of $\Psi$
and $\delta\phi$. This solution is the decaying solution, therefore, 
one can set $\delta X=0$. 
%\footnote{
%We considered two cases of the initial condition for $\dot{\delta \phi}$. 
%One is $\dot{\delta \phi}=0$, while the
%other is $\dot{\delta \phi}=\dot\phi\Psi$ so that $\delta X=0$. 
%However, the cosmological perturbations become the same   
%soon after the beginning of the evolution. }
In the case $\delta X=0$, the galileon field perturbations in 
the right hand sides of eqs.~(\ref{fullPertEij}) and (\ref{fullPertE0i})
disappear, and the equations reduce to those of the $\Lambda{\rm CDM}$ model.
The evolution equations of the matter perturbations are not altered
unless the gravitational potentials are altered. 
We are considering the initial conditions that is the same as the 
$\Lambda$CDM model.
These suggest that the linear perturbations of the kinetic gravity braiding 
model for $n=\infty$ are equivalent to those of the 
$\Lambda$CDM model. 
In our numerical computation, we have confirmed that the cosmological 
perturbations of the kinetic braiding model with large $n$ 
approach the result of the $\Lambda$CDM model (see also subsection
\ref{FullNC}).

%Therefore, 
%the gravitational potential $\Psi$ and the velocity $v$ respectively
%share the same history as the $\Lambda {\rm CDM}$ model. 
%From Eq.~(\ref{fullDensityEvolution}), the evolution of the 
%perturbated matter density is also same. Thus both the background 
%and cosmological perturbation is identical to the $\Lambda {\rm CDM}$ model. 

This property that our model reduces to the cosmological constant model
can be understood in the Lagrangian 
(\ref{Lagrangian}) with eqs.~(\ref{K}) and (\ref{G}).
In the case, $\dot{\phi}={\rm constant}$ and $\delta X=0$,
$K(X)$ and $G(X)$ in the Lagrangian are constants, then
the term $G\square\phi$ can be written as the total derivative,
which can be dropped effectively when deriving the equation of
motion. The term $K(X)$ plays a role of the cosmological 
constant, and the kinetic braiding model reduces to the 
cosmological constant model for $n=\infty$.
This feature could be related to the fact that 
the sound speed of perturbations of 
the galileon field for $n=\infty$ is zero over 
the whole history of the universe. 
The perturbations of the galileon field can not propagate and 
does not induce an additional effect on gravity. 

The feature that the linear perturbations of the kinetic 
braiding model approaches those of the cosmological constant model 
appears for very large $n$, e.g., $n\simgt1000$, as will be 
shown in the next subsection. Such a behavior of large $n$ 
model is different from that of smaller $n$ model, e.g., 
$n\simlt10$. 
For the case $n\simlt10$, the quasi-static approximation can 
be adopted for the galileon field perturbations on the 
subhorizon scales, while it is not valid for $n\simgt1000$. 
This comes from the behavior of the sound speed of the galileon 
field perturbations. 
In the k-essence model with a zero sound speed, the scalar
field perturbations cluster, and enhance the cosmological 
perturbations \cite{Creminelli09112701}, which is a contrast 
to our kinetic braiding model on the attractor solution (\ref{Att}).

%%%%%%%%%%%%%%%%%%%%%%%%%%%%%%%%%%%%%%%%%%%%%%%%%%%%%%%%%%%%%
\subsection{Numerical Calculation}
\label{FullNC}
\begin{figure}[t]
  \begin{tabular}{cc}
    \begin{minipage}{0.5\textwidth}
      \begin{center}
        \includegraphics[scale=1.25]{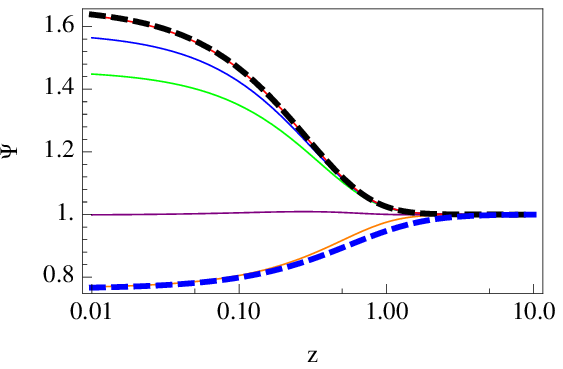}
      \end{center}
    \end{minipage}
    \begin{minipage}{0.5\textwidth}
      \begin{center}
        \includegraphics[scale=1.25]{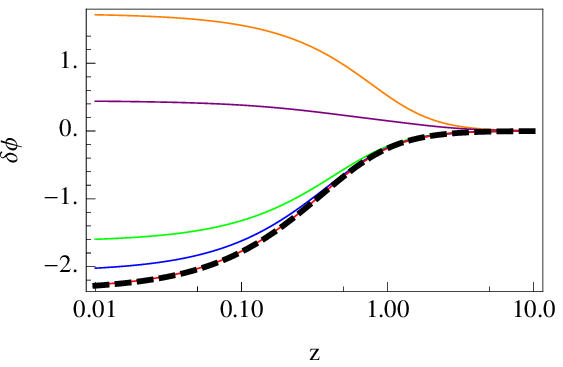}
      \end{center}
    \end{minipage}
    \\
    \begin{minipage}{0.5\textwidth}
      \begin{center}
        \includegraphics[scale=1.25]{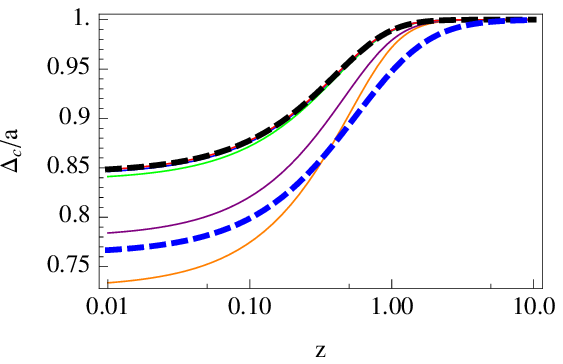}% 
      \end{center}
    \end{minipage}
    \begin{minipage}{0.5\textwidth}
      \begin{center}
        \includegraphics[scale=1.1]{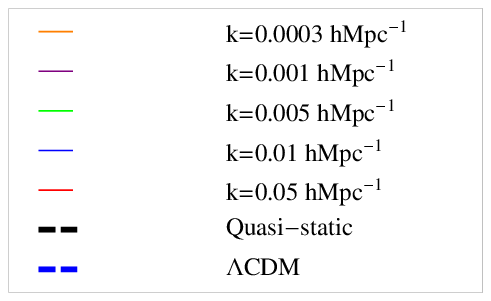}
      \end{center}
    \end{minipage}
  \end{tabular}
  \caption{The gravitational potential $\Psi$ (upper left panel),
    the perturbation of the galileon filed (upper right panel),
    and the growth factor divided by scale factor (lower panel)
    as a function of redshift for the $\Lambda$CDM model 
    (thick dotted curve) for the kinetic braiding model of $n=1$ 
    with $k=0.0003 h{\rm Mpc^{-1}}$, $0.001 h{\rm Mpc^{-1}}$, 
    $0.005 h{\rm Mpc^{-1}}$, $0.01 h{\rm Mpc^{-1}}$, 
    $0.05 h{\rm Mpc^{-1}}$, and the case of the quasi-static 
    approximation (see subsection \ref{QSA}),  respectively. 
    The gravitational potential $\Psi$ is normalized to unity 
    at the initial time.
    \label{fig_n1}}
  %%%%%%%%%%%%% n=10 %%%%%%%%%%%%%%%
  % \begin{figure}[]
  \begin{tabular}{cc}
    \begin{minipage}{0.5\textwidth}
      \begin{center}
        \includegraphics[scale=1.25]{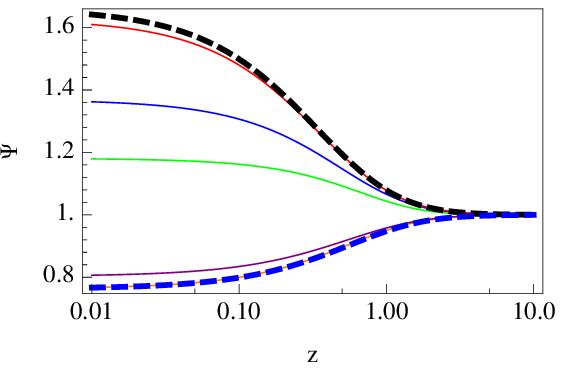}
      \end{center}
    \end{minipage}
    \begin{minipage}{0.5\textwidth}
      \begin{center}
        \includegraphics[scale=1.25]{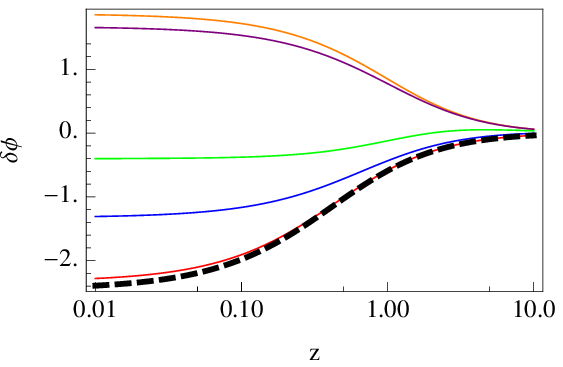}
      \end{center}
    \end{minipage}
    \\
    \begin{minipage}{0.5\textwidth}
      \begin{center}
        \includegraphics[scale=1.25]{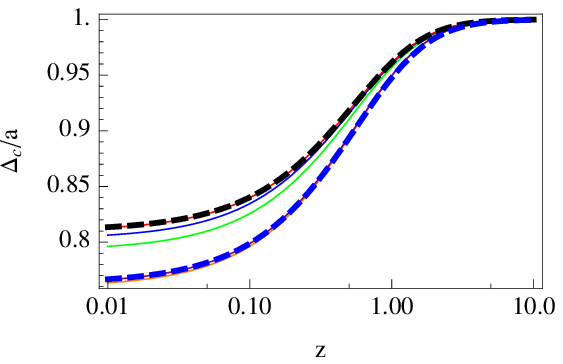}% 
      \end{center}
    \end{minipage}
    \begin{minipage}{0.5\textwidth}
      \begin{center}
        \includegraphics[scale=1.1]{legend.eps}
      \end{center}
    \end{minipage}
  \end{tabular}
  \caption{Same figure as figure~\ref{fig_n1} but with $n=10$.
    \label{fig_n10}}
\end{figure}

\begin{figure}[]
  \begin{tabular}{cc}
   \begin{minipage}{0.5\textwidth}
    \begin{center}
     \includegraphics[scale=1.25]{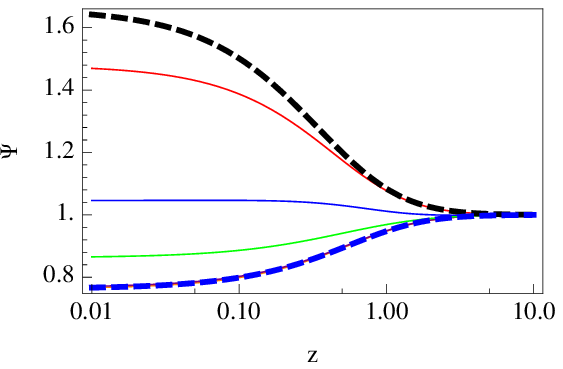}
    \end{center}
   \end{minipage}
   \begin{minipage}{0.5\textwidth}
    \begin{center}
     \includegraphics[scale=1.25]{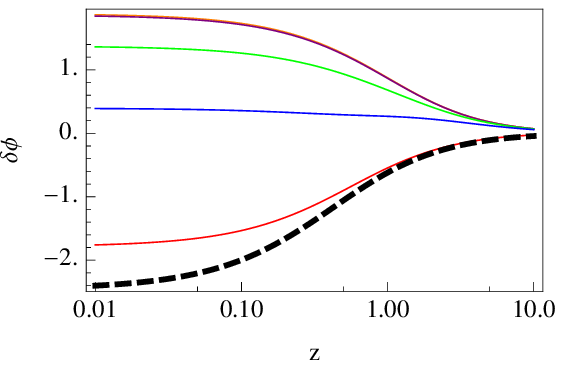}
    \end{center}
   \end{minipage}
\\
   \begin{minipage}{0.5\textwidth}
    \begin{center}
     \includegraphics[scale=1.25]{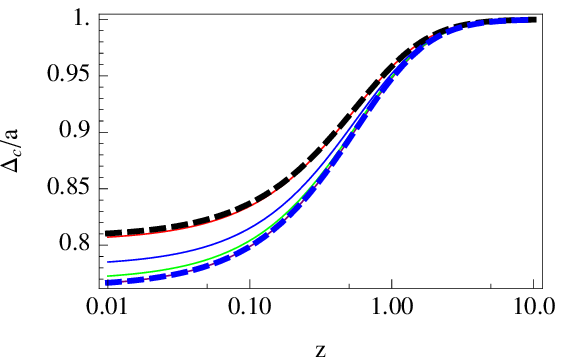}% 
    \end{center}
   \end{minipage}
   \begin{minipage}{0.5\textwidth}
    \begin{center}
     \includegraphics[scale=1.1]{legend.eps}
    \end{center}
   \end{minipage}
  \end{tabular}
  \caption{Same figure as figure~\ref{fig_n1} but with $n=100$.
  \label{fig_n100}}
\end{figure}

\begin{figure}[]
  \begin{tabular}{cc}
   \begin{minipage}{0.5\textwidth}
    \begin{center}
     \includegraphics[scale=1.25]{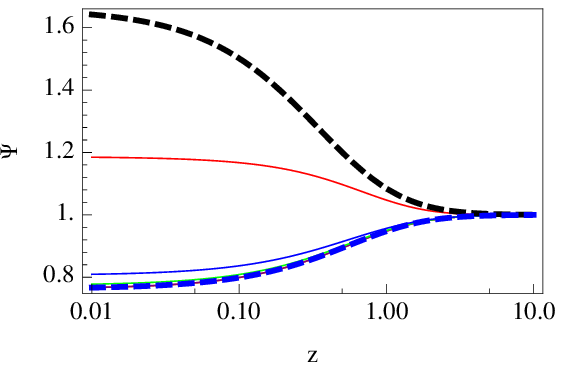}
    \end{center}
   \end{minipage}
   \begin{minipage}{0.5\textwidth}
    \begin{center}
     \includegraphics[scale=1.25]{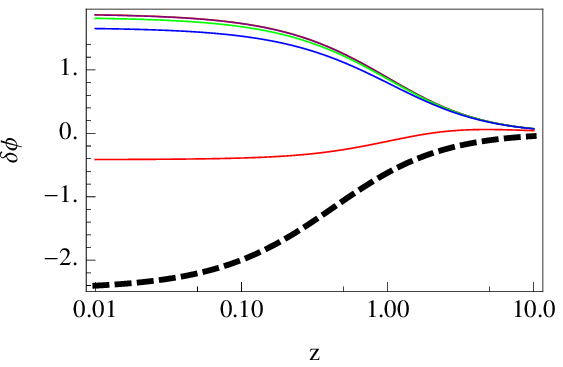}
    \end{center}
   \end{minipage}
\\
   \begin{minipage}{0.5\textwidth}
    \begin{center}
  \includegraphics[scale=1.25]{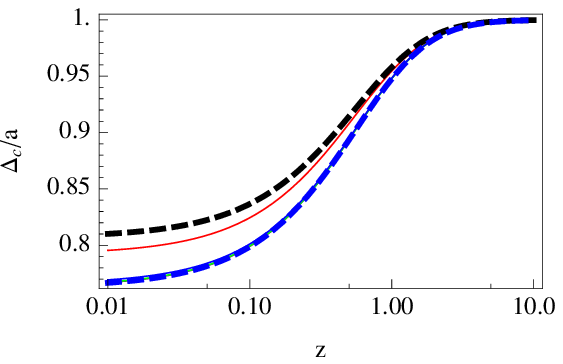}% 
    \end{center}
   \end{minipage}
   \begin{minipage}{0.5\textwidth}
    \begin{center}
     \includegraphics[scale=1.1]{legend.eps}
    \end{center}
   \end{minipage}
  \end{tabular}
  \caption{Same figure as figure~\ref{fig_n1} but with $n=1000$.
  \label{fig_n1000}}
\end{figure}

%%%%%%%%%%%%%%%%%%%%%%%%%%%%%%%%%%%%%%%%%%%%%%%%%%%%%%%%%%%%%
In this subsection, we show the evolution of the perturbations 
solved numerically. 
In figure~\ref{fig_n1}, we plotted the evolutions of the gravitational 
potential $\Psi$ (upper left panel), the perturbations of the galileon 
field $\delta\phi$ (upper right panel), and the density perturbations 
$\Delta_c$ (lower panel), for the kinetic braiding model with
$n=1$ with adopting different wavenumbers labeled there and 
for the $\Lambda{\rm CDM}$ 
model. Figs.~\ref{fig_n10},~\ref{fig_n100} and \ref{fig_n1000} are the
same as figure~\ref{fig_n1} but with $n=10$, $n=100$ and $1000$,
respectively. 
%We find that the quasi-static approximation is valid for the wavelength 
%$k\simgt0.01 h{\rm Mpc^{-1}}$ for $n\simlt 10$. However, 
%the quasi-static approximation no longer works for $n\simgt100$.
We can see that the perturbations approach those of the 
$\Lambda{\rm CDM}$ model as $n$ becomes large, as we discussed 
in subsection \ref{secInf}.
For $n\simgt 1000$, the perturbations $\Psi$ and $\Delta_c$ 
with $k\simlt 0.01h{\rm Mpc}^{-1}$ are almost the same as 
those of the $\Lambda$CDM model, as one can see from figure 
\ref{fig_n1000}.
Since the sound speed of the perturbed galileon field become 
smaller as $n$ increases, longer wavelength perturbations 
with larger $n$ are not affected by the galileon field. 
Thus, the longer wavelength perturbations approach the 
$\Lambda{\rm CDM}$ model quickly as $n$ increases. 

\begin{figure}[]
  \begin{tabular}{cc}
   \begin{minipage}{0.5\textwidth}
    \begin{center}
     \includegraphics[scale=1.25]{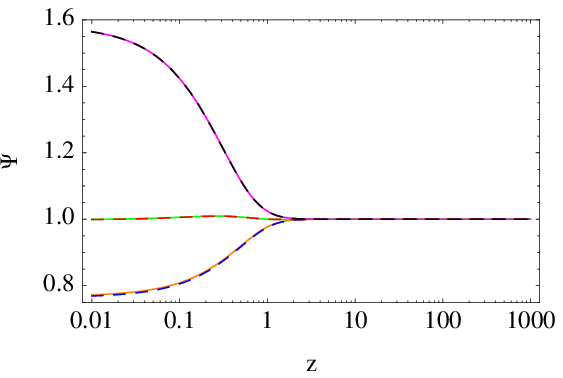}
    \end{center}
   \end{minipage}
   \begin{minipage}{0.5\textwidth}
    \begin{center}
     \includegraphics[scale=1.25]{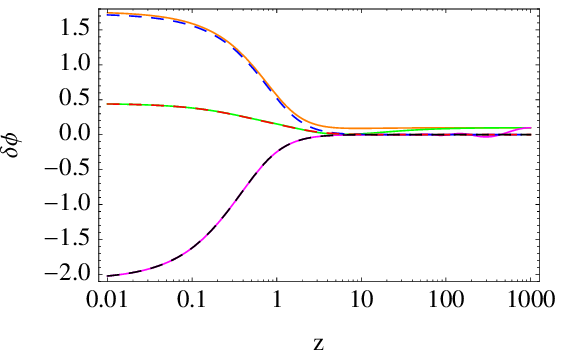}
    \end{center}
   \end{minipage}
\\
   \begin{minipage}{0.5\textwidth}
    \begin{center}
  \includegraphics[scale=1.25]{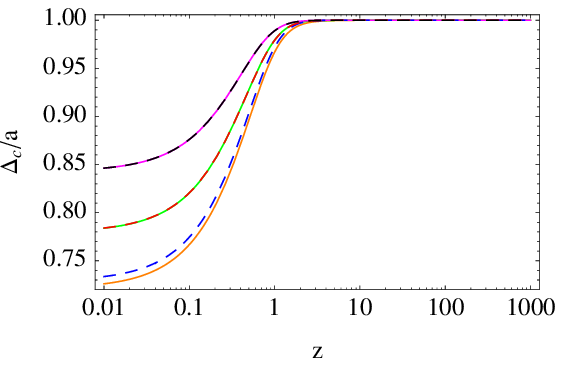}% 
    \end{center}
   \end{minipage}
   \begin{minipage}{0.5\textwidth}
    \begin{center}
     \includegraphics[scale=1.1]{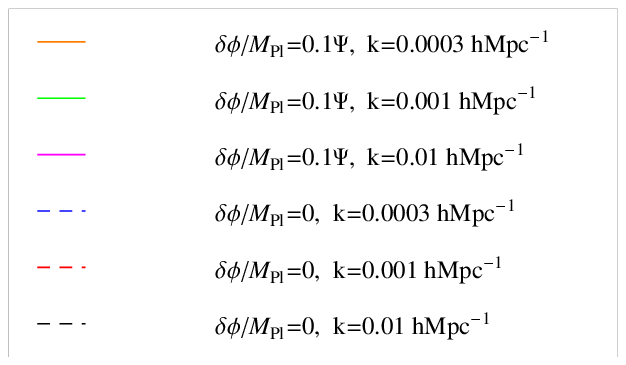}
    \end{center}
   \end{minipage}
  \end{tabular}
  \caption{
     The gravitational potential $\Psi$ (upper left panel),
     the perturbation of the galileon filed (upper right panel),
     and the growth factor divided by scale factor (lower panel)
     as a function of redshift for the kinetic braiding model of $n=1$ 
     for $k=0.0003 h{\rm Mpc^{-1}}$, $0.001 h{\rm Mpc^{-1}}$, 
     $0.01 h{\rm Mpc^{-1}}$ with different initial conditions 
     $\delta\phi_i/{M_{\rm Pl}}=0.1\Psi_i$ (solid curve) and 
     $\delta\phi_i/{M_{\rm Pl}}=0$ (dashed curve), respectively.
     We adopt the initial condition $\dot{\delta\phi}_i=0$. 
  \label{fig_ini0}}
\end{figure}

\begin{figure}[]
  \begin{tabular}{cc}
   \begin{minipage}{0.5\textwidth}
    \begin{center}
     \includegraphics[scale=1.25]{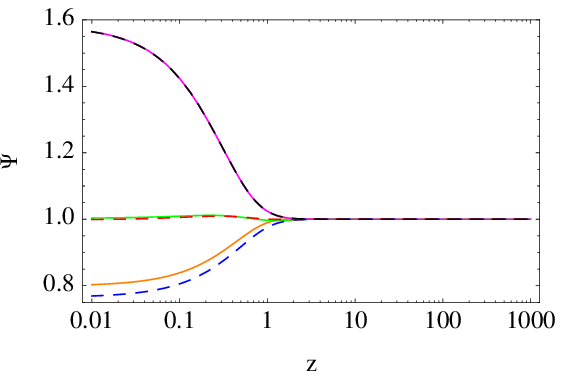}
    \end{center}
   \end{minipage}
   \begin{minipage}{0.5\textwidth}
    \begin{center}
     \includegraphics[scale=1.25]{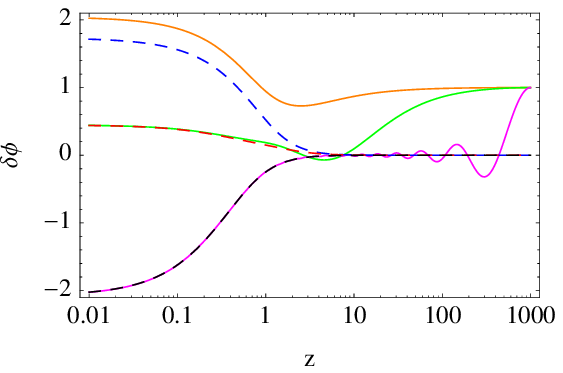}
    \end{center}
   \end{minipage}
\\
   \begin{minipage}{0.5\textwidth}
    \begin{center}
  \includegraphics[scale=1.25]{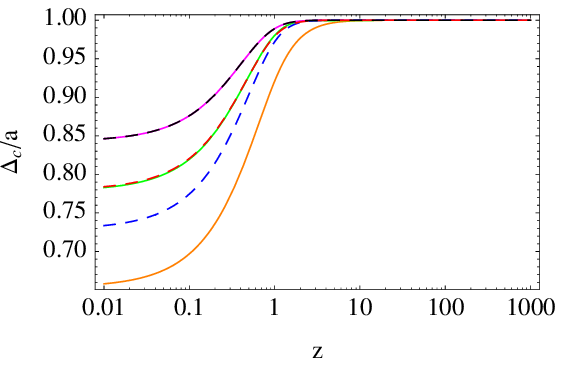}% 
    \end{center}
   \end{minipage}
   \begin{minipage}{0.5\textwidth}
    \begin{center}
     \includegraphics[scale=1.1]{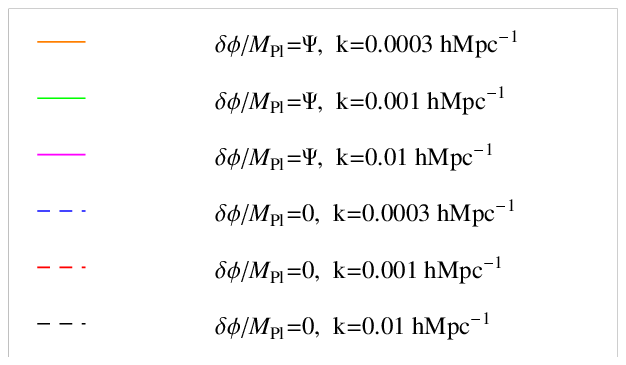}
    \end{center}
   \end{minipage}
  \end{tabular}
  \caption{Same figure as figure~\ref{fig_ini0} but with 
$\delta\phi_i/{M_{\rm Pl}}=\Psi_i$.
  \label{fig_ini1}}
\end{figure}

\begin{figure}[]
  \begin{tabular}{cc}
   \begin{minipage}{0.5\textwidth}
    \begin{center}
     \includegraphics[scale=1.25]{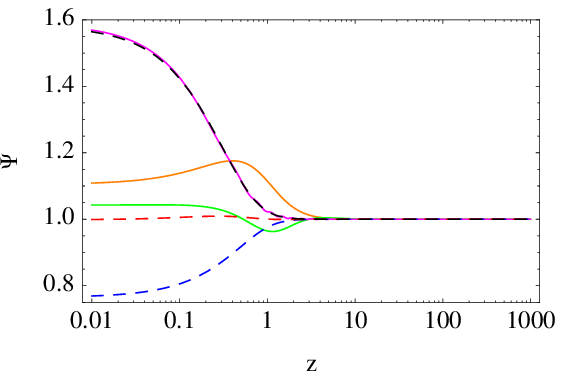}
    \end{center}
   \end{minipage}
   \begin{minipage}{0.5\textwidth}
    \begin{center}
     \includegraphics[scale=1.25]{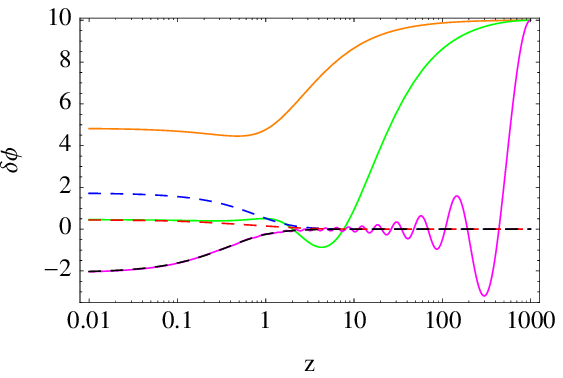}
    \end{center}
   \end{minipage}
\\
   \begin{minipage}{0.5\textwidth}
    \begin{center}
  \includegraphics[scale=1.25]{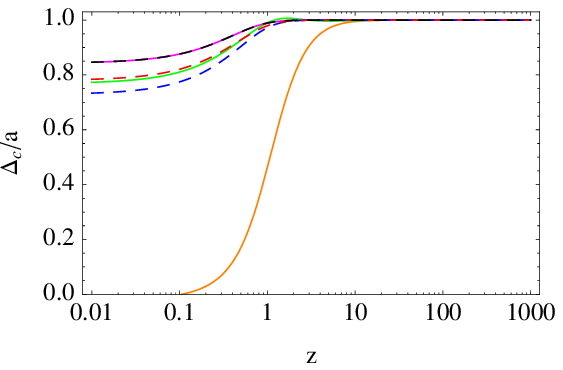}% 
    \end{center}
   \end{minipage}
   \begin{minipage}{0.5\textwidth}
    \begin{center}
     \includegraphics[scale=1.1]{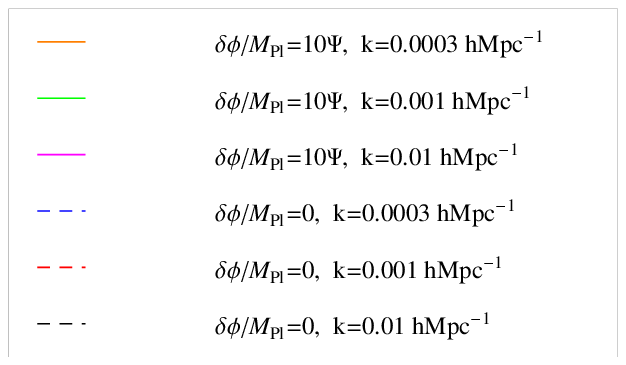}
    \end{center}
   \end{minipage}
  \end{tabular}
  \caption{Same figure as figure~\ref{fig_ini0} but with 
$\delta\phi_i/{M_{\rm Pl}}=10\Psi_i$.
  \label{fig_ini10}}
\end{figure}

Let us discuss about the choice of the initial condition
of the numerical computation. For the galileon field 
perturbations, the above results adopted the initial 
conditions $\delta\phi_i=0$ and $\dot{\delta\phi}_i=0$, 
where the subscript $i$ denotes the initial values at $a_i=10^{-3}$.
In the Galileon scalar-tensor models \cite{ELCP}, it has been 
pointed out that a different choice of the
initial condition might give rise to an oscillating behavior 
in $\Psi$, $\delta\phi$, and $\Delta_c$.
Figs.~\ref{fig_ini0}, \ref{fig_ini1} and \ref{fig_ini10} compares the 
gravitational potential $\Psi$, the perturbation of the galileon 
field $\delta\phi$, and the growth factor $\Delta_c$, as a function 
of the scale factor, for the kinetic braiding model $n=1$ with the 
different initial conditions $\delta\phi_i/M_{\rm Pl}=0$, 
$\delta\phi_i/M_{\rm Pl}=0.1\Psi_i$, 
$\delta\phi_i/M_{\rm Pl}=\Psi_i$ and $\delta\phi_i/M_{\rm Pl}=10\Psi_i$, 
respectively, where the other parameters are the same as those of 
Fig.~\ref{fig_n1}. 
%Figs.~\ref{fig_ini0}, \ref{fig_ini1}  and \ref{fig_ini10} show
%that the cosmological perturbations $\Psi$ and $\Delta_C$ with 
%the initial condition $\delta\phi_i=\Psi_i$ and 
%$\delta\phi_i=10\Psi_i$ approach 
%that with the initial condition $\delta\phi=0$. 
The behavior of the cosmological perturbations 
$\Psi$ and $\Delta_c$ are almost same for 
%$a\simgt 0.1$ and 
$k\simgt0.001{h{\rm Mpc}^{-1}}$ as long as 
$|\delta\phi_i|/M_{\rm Pl}\simlt \Psi_i$. 
The difference becomes smaller as the initial value 
$|\delta\phi_i|/M_{\rm Pl}$ is smaller. 
Even for the cases $\delta\phi_i/M_{\rm Pl}=\Psi_i$ and
$\delta\phi_i/M_{\rm Pl}=10\Psi_i$, the 
the difference appears only for the large scales 
$k\simlt0.0003{h{\rm Mpc}^{-1}}$. 
We also checked that the different choice of $\dot{\delta\phi}_i$
does not alter this conclusion as long as 
%$\dot{\delta\phi}_i/\dot\phi_i
$|\dot{\delta\phi}_i|/H_iM_{\rm Pl}
\simlt \Psi_i$, where $H_i$ is the Hubble parameter at the 
initial time $a_i=10^{-3}$.
Therefore, the result is not very sensitive to the choice 
of the initial condition within a suitable range of it. 

If we assume that the galileon field undergoes an inflationary period
at the early stage of the universe, fluctuations of order 
$\delta \phi_i \sim H_{\rm inf}/2\pi$ could be set, 
where $H_{\rm inf}$ is the Hubble scale determined by the energy scale 
of the inflation.
The condition $|\delta\phi_i|/M_{\rm Pl}\simlt \Psi_i$ yields
$H_{\rm inf}/(2\pi M_{\rm Pl})\simlt \Psi_i\sim10^{-5}$, 
which might be satisfied as long as the energy scale 
of the inflation is not so high.

%$\delta\phi_i=0$ and $\dot{\delta\phi}_i=0$
%Therefore, we can safely choose the initial conditions 
%$\delta\phi_i=0$ and $\dot{\delta\phi}_i=0$.

For the model with larger $n$, the quasi-static approximation 
for the galileon field perturbation fails. 
We can estimate the condition for the validity of the quasi-static 
approximation, as follows. From the equation of motion obtained from
(\ref{stabilityAction}), the condition is given by 
$\kappa \ddot{\delta\phi} \ll \beta k^2 \delta\phi /a^2$.
Using the relations $\dot{\delta\phi}\sim H\delta\phi$ and 
$\ddot{\delta\phi} \sim H^2\delta\phi$, the quasi-static approximation 
is valid as long as $k^2c_s^2/a^2 \gg H^2$.
We find that the quasi-static approximation is valid for the wavenumber 
$k\simgt0.01 h{\rm Mpc^{-1}}$ for $n\simlt 10$ from numerical results. 
However, the quasi-static approximation no longer works for $n\simgt100$.
This is because the sound speed of perturbations of the galileon field 
approaches zero for very large $n$.

%%%%%%%%%%%%% n=1 %%%%%%%%%%%%%%%

\vspace{2cm}

%%%%%%%%%%%%% n=100 %%%%%%%%%%%%%%%

%%%%%%%%%%%%% n=1000 %%%%%%%%%%%%%%%

%%%%%%%%%%%%% n=1, deltaphi=0.1 %%%%%%%%%%%%%%%
%%%%%%%%%%%%% n=1, deltaphi=1 %%%%%%%%%%%%%%%
%%%%%%%%%%%%% n=1, deltaphi=10 %%%%%%%%%%%%%%%

%%%%%%%%%%%%%%%%%%%%%%%%%%
\subsection{Quasi-Static Approximation}
\label{QSA}
\begin{figure}[]
  \begin{tabular}{cc}
   \begin{minipage}{0.5\textwidth}
    \begin{center}
     \includegraphics[scale=1.4]{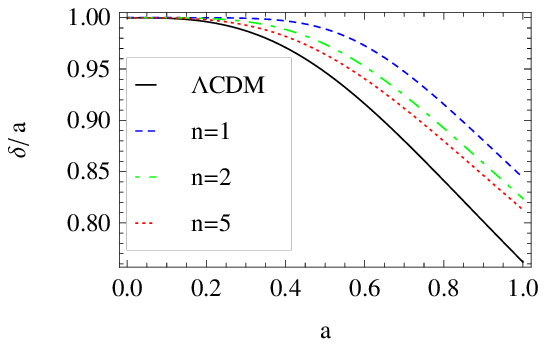}
    \end{center}
   \end{minipage}
   \begin{minipage}{0.5\textwidth}
    \begin{center}
     \includegraphics[scale=1.4]{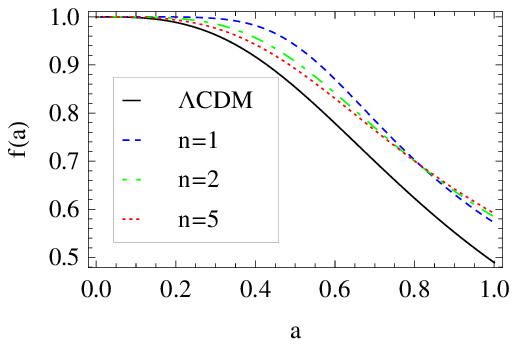}
    \end{center}
   \end{minipage}
  \end{tabular}
  \caption{Left panel: The growth factor divided by scale factor 
as a function of scale factor for the $\Lambda$CDM model (solid curve) and
the kinetic braiding model $n=1$ (dashed curve), 
  $n=2$ (dash-dotted curve), and $n=5$ (dotted curve), respectively. 
Right panel: The linear growth rate as a function of scale factor.
  \label{fig4}}
\end{figure}

%%%%%%%%%%%%%%%%%%%%%%%%%%
Hereafter, we consider the model with $n\simlt 10$,
in which small-scale perturbation equations can be expressed in 
simple forms by using the quasi-static and 
the sub-horizon approximation. 
Neglecting the time-derivative terms of perturbations and assuming 
that the wavelength of perturbation is well inside the Hubble 
horizon, 
$\mathcal{O}(k^2c_s^2/a^2) \gg \mathcal{O}(H^2)$, 
%$\mathcal{O}(\nabla^2\Phi/a^2) \gg \mathcal{O}(H^2\Phi)$, 
the evolution equation for the matter overdensity $\delta$ 
in linear theory reduces to
\begin{eqnarray}
  \ddot{\delta}+2H\dot{\delta} \simeq \frac{\nabla^2}{a^2}\Psi,
\label{evodelta}
\end{eqnarray}
where we used the fact $\delta=\delta\rho/\rho\simeq \Delta_c$
(see also appendix A).
{}From eqs.~(\ref{fullPertEij}) and (\ref{fullPertEtl}), we have
\begin{eqnarray}
  &&-\frac{\nabla^2}{a^2}\Psi \simeq -4\pi G \rho \delta
+ 4\pi G G_X\dot\phi^2\phi\frac{\nabla^2}{a^2}\varphi,
\label{EinsteinP2}
\end{eqnarray}
where we introduced $\varphi(x,t)=\delta\phi(x,t)/\phi(t)$.
{}From eq.~(\ref{fullPertField}), we have
\begin{eqnarray}
  \beta(a) \frac{\nabla^2}{a^2}\varphi \simeq -4\pi G \frac{G_X\dot\phi^2}{\phi}\rho\delta,
  \label{LinearFieldEq}
\end{eqnarray}
for the galileon field perturbation. 
Combining (\ref{EinsteinP2}) and (\ref{LinearFieldEq}), 
we obtain the modified Poisson equation,
\begin{eqnarray}
  \frac{\nabla^2}{a^2}\Psi \simeq 4\pi G_{\rm eff}\rho \delta,
\label{Poisson}
\end{eqnarray}
where the effective gravitational constant is given by 
\begin{eqnarray}
 G_{\mathrm{eff}}&=&G\left[1+4 \pi G \frac{G_X^2\dot\phi^4}{\beta(a)}\right]
\nonumber
\\
 &=&G\frac{2n+3n\Omega_{\rm m}(a)-\Omega_{\rm m}(a)}{\Omega_{\rm m}(a)(5n-\Omega_{\rm m}(a))}.
  \label{geffomega}
\end{eqnarray}
{Here, in the second line, we used the attractor condition (\ref{Att}).}
Then, the evolution equation for the matter overdensity in linear theory 
can be written as 
\begin{eqnarray}
  \ddot{\delta}+2H\dot{\delta} \simeq 4\pi G_{\mathrm{eff}}\rho\delta.
\label{DensityPerturbation}
\end{eqnarray}
The crucial difference between the kinetic braiding model and the 
scalar-tensor galileon theory \cite{SAUGC,ELCP,CEGH} is 
the absence of an effective anisotropic stress in the 
right-hand-side of eq.~(\ref{fullPertEtl}). 
The effective gravitational constant $G_{\mathrm{eff}}$ is nearly 
equal to $G$ at early times and becomes larger than $G$ at late times.
(See the dashed curve in the right panel of figure~\ref{SCollaseFig}). 
The enhancement of the effective gravitational constant 
leads to an enhancement of the growth factor of matter density perturbations. 
Although the background evolution  
for large $n$ approaches the $\Lambda\mathrm{CDM}$ model, 
the growth history is different due to the time-dependent 
effective gravitational constant. 
The left panel of figure~\ref{fig4} compares the evolution of 
the growth factor divided by the scale factor, $\delta/a$, which 
is normalized as $1$ at early stage of the evolution.
The growth factor of the kinetic braiding model is larger than 
that of the $\Lambda \mathrm{CDM}$ model throughout its growth history. 
In the right panel of figure~\ref{fig4}, the linear growth rate 
$f(a)=d\ln \delta / d\ln a$ is plotted.

\subsection{Analytic Formula for Growth Index}

The growth rate can be parametrized by the 
growth index $\gamma(a)$ defined by $f(a)=\Omega_{\rm m} (a)^{\gamma(a)}$. 
It has been known that the growth rate of the $\mathrm{\Lambda CDM}$ model
is well approximated by the nearly constant value $\gamma=6/11$.
This parametrization was first introduced in \cite{Peebles} 
and has the potential to distinguish dark energy models and 
modified gravity models \cite{Linder}. 
In this section, we derive the analytic formula of the growth index 
for the kinetic braiding model. 
As we will show below, the behavior of the growth index is
unique in the kinetic braiding model, which will be important 
to distinguish between the kinetic braiding model and the
$ \Lambda \mathrm{CDM}$ model when $n$ is not infinite.

\begin{figure}[t]
\begin{center}
\includegraphics[scale=1.4]{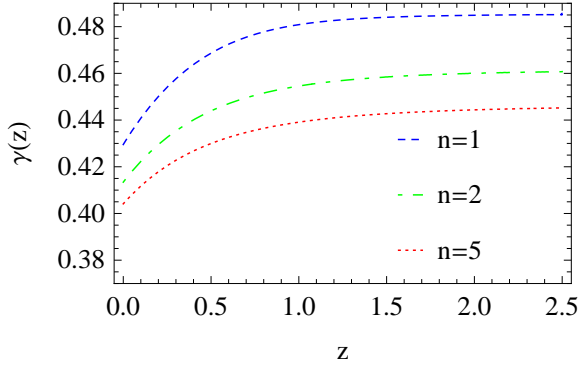}
\caption{The growth index of the kinetic braiding model 
as a function of redshift for $n=1$ (dashed curve), 
$n=2$ (dash-dotted curve), and $n=5$ (dotted curve), 
from the top to bottom, respectively.
\label{fig5} {}}
\end{center}
\end{figure}

\begin{figure}[h]
  \begin{tabular}{cc}
   \begin{minipage}{0.5\textwidth}
    \begin{center}
     \includegraphics[scale=1.4]{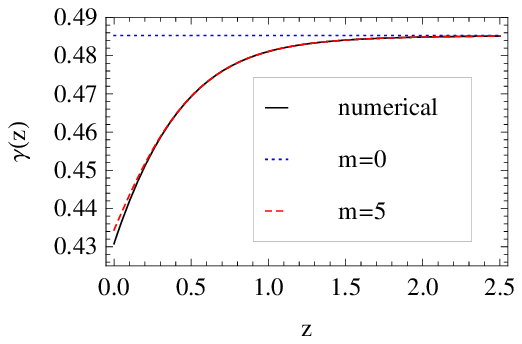}
    \end{center}
   \end{minipage}
   \begin{minipage}{0.5\textwidth}
    \begin{center}
     \includegraphics[scale=1.4]{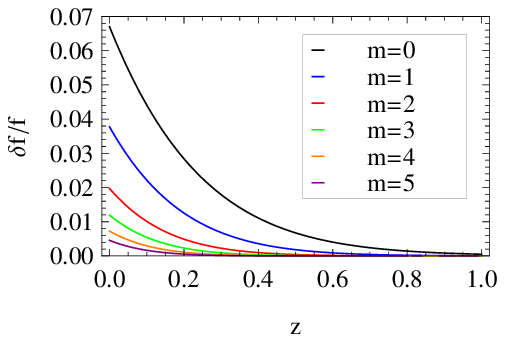}
    \end{center}
   \end{minipage}
  \end{tabular}
  \caption{Comparison between the numerical result and the approximate
formula of (\ref{Pgamma}) for the model $n=1$.
Left panel is the growth index as a function of redshift.
The solid curve is the exact result from numerically solving the 
evolution equation, but the other two curves are from the 
formula in the expansion (\ref{Pgamma}) with $m=1$ (dotted curve) 
and $m=5$ (dashed curve), respectively.
Right panel is the relative error $\delta f/f$ for the case when the
maximum value $m$ is set as $0,~1,~2,~3,~4,$ and $5$, respectively.
  \label{fig6} {}}
\end{figure}

The matter evolution equation (\ref{DensityPerturbation}) can be 
written in terms of the growth rate $f(a)=d\ln \delta / d\ln a$ as 
\begin{eqnarray}
  {df\over d\ln a}+f^2+\left(2+\frac{\dot{H}}{H^2}\right)f-\frac{3}{2}\frac{G_{\mathrm{eff}}}{G}\Omega_{\rm m}(a)=0.
\label{DensityPerturbationf}
\end{eqnarray}
By using the background equation
and (\ref{Att})
, we have
\begin{eqnarray}
  \frac{\dot{H}}{H^2}&=&-\frac{3(2n-1)\Omega_{\rm m}(a)}{2(2n-\Omega_{\rm m}(a))},
  \label{hhdot}
\\
  \frac{d\Omega_{\rm m}(a)}{d\ln a}&=&-\frac{6n\Omega_{\rm m}(a)(1-\Omega_{\rm m}(a))}{2n-\Omega_{\rm m}(a)}.
  \label{omegadash}
\end{eqnarray}
Using eqs.~(\ref{geffomega}), (\ref{hhdot}) and (\ref{omegadash}), 
eq.~(\ref{DensityPerturbationf}) can be written as 
\begin{eqnarray}
  &&-\frac{6n\Omega_{\rm m}^2(a) (1-\Omega_{\rm m}(a))}{2n-\Omega_{\rm m}(a)}
\left\{\frac{d\gamma(a)}{d\Omega_{\rm m}(a)}\ln \Omega_{\rm m}(a)
+{\gamma(a)\over \Omega_{\rm m}(a)}\right\}
+\Omega_{\rm m}(a)^{2\gamma(a)}
\nonumber
\\
&&~~
+\frac{8n-6n\Omega_{\rm m}(a)-\Omega_{\rm m}(a)}{2(2n-\Omega_{\rm m}(a))}\Omega_{\rm m}(a)^{\gamma(a)}
%\nonumber
%\\
%&&~~
-\frac{3(2n+3n\Omega_{\rm m}(a)-\Omega_{\rm m}(a))}{2(5-\Omega_{\rm m}(a))}=0.
\label{GammaEq}
\end{eqnarray}
We assume that the growth index can be approximated in the 
following expanded form,
\begin{eqnarray}
  \gamma(a)\simeq\sum_{j=0}^{m}\zeta_j(1-\Omega_{\rm m}(a))^j.
  \label{Pgamma}
\end{eqnarray}
Then, we expand eq.~(\ref{GammaEq}) around $\Omega_{\rm m}(a) =1$ and 
we have the asymptotic value of the growth index at high redshift,
\begin{eqnarray}
  \zeta_0=\frac{3n(16n-5)}{110n^2-47n+5}.
\label{zeta0}
\end{eqnarray}
The higher order coefficients can be easily found.
%For example,
By way of example,
table \ref{table1} summarizes the coefficients for the 
model $n=1,~2,~3,~4,$ and $5$.
The redshift evolution of $\gamma(a)$
is important in the kinetic braiding model. 
This is shown in figure~\ref{fig5}, which plots 
the growth index $\gamma(a)$, by solving eq.~(\ref{DensityPerturbationf}) 
numerically, as a function of redshift, for $n=1,~2,$ and $5$,
respectively. 
Figure~\ref{fig6} compares the numerical result and the approximate
formula of (\ref{Pgamma}) of the model $n=1$ as a function of redshift.
The left panel of figure~\ref{fig6}
plots the numerical result and the approximate formula with 
$m=0$ and $m=5$, respectively, where $m$ is the maximum 
index in (\ref{Pgamma}).
The right panel shows the 
relative error $(f-\Omega_{\rm m}(a)^{\gamma(a)})/f$ for $m=0,~1,~2,~3,~4,$ and 
$5$. 
The approximate formula (\ref{Pgamma}) with $m=5$ 
reproduces the exact growth rate $f$ to better than 
$0.46\%$, and the growth factor can be obtained 
in a highly accurate manner, by $D_1(a)=a\exp\left[\int_0^a d\ln a' \left\{\Omega_{\rm m}(a')^{\gamma}-1 \right\}\right]$.
For the other case in table \ref{table1}, 
we have the similar result and confirmed that the growth 
factor can be well approximated by the formula up to $m=5$. 


%%%%%%%%%%%%%%%%%%%%%%%%%%%%%%%%%%%%%%%%%%%%%%%%%%%%%%%%%%%%%%%%%%%%%%%%
\begin{table}[htbp]
 \begin{center}
  \begin{tabular}{l|lllll}
   \hline
   ~~~~~~ &~ ~$n=1$~~~~  & ~$n=2$~~~~ & ~$n=3$~~~~ & ~$n=4$~~~~ & ~$n=5$
~~~~\\
   \hline
   ~~$\zeta_0$ &~  0.48529   & 0.46154 &  0.45316 &   0.44895 &    0.44643 \\
   ~~$\zeta_1$ &~ -0.03373  & -0.03153 & -0.02922 &  -0.02791 &   -0.02708 \\
   ~~$\zeta_2$ &~ -0.02814  & -0.02037 & -0.01876 &  -0.01805 &   -0.01764 \\
   ~~$\zeta_3$ &~ -0.01667  & -0.01470 & -0.01393 &  -0.01355 &   -0.01333 \\
   ~~$\zeta_4$ &~ -0.01367  & -0.01174 & -0.01119 &  -0.01093 &   -0.01078 \\
   ~~$\zeta_5$ &~ -0.01106  & -0.00976 & -0.00936 &  -0.00916 &   -0.00905 \\
%   ~~$\zeta_6$ &~ -0.009377  & -0.008345 & -0.008030 &  -0.007877 &   -0.007786% \\
   \hline
  \end{tabular}
  \caption{Coefficients $\zeta_j$ for the model of $n=1,~2,~3,~4,$ and $
5$. \label{table1}}
 \end{center}
\end{table}
%%%%%%%%%%%%%%%%%%%%%%%%%%%%%%%%%%%%%%%%%%%%%%%%%%%%%%%%%%%%%%%%%%%%%%%%

On very small scales, the non-linear terms of  $\varphi$ 
become more important than metric perturbations
in the case of small $n$. 
%as long as $n<10$.
The field equation for the galileon field including 
the non-linear terms is given by
\begin{eqnarray}
  \beta(a)\frac{\nabla^2}{a^2}\varphi+\frac{G_X \phi}{a^4}\left[(\nabla_i\nabla_j\varphi)(\nabla^i\nabla^j\varphi)
-(\nabla^2\varphi)^2\right]=-4\pi G \frac{G_X\dot\phi^2}{\phi}\rho\delta,
\label{NonlinearFieldEq}
\end{eqnarray}
which holds for the general Lagrangian (\ref{Lagrangian}) 
as long as the quasi-static approximation works. 
The equations for the metric perturbation remain the same as those
of the linear perturbation theory.
{}From eq.~(\ref{NonlinearFieldEq}),
the non-linear terms in the galileon field equation 
become dominant when $\varphi > \beta a^2/k^2 G_X \phi$. 
{}From eq.~(\ref{LinearFieldEq}), we may estimate
$\varphi \sim a^2 H^2 G_X \dot{\phi}^2 \delta / k^2 \beta\phi$, by using the 
relation, $H^2 \simeq \rho_m/3M_{Pl}^2 $ . 
Combining these relations, the non-linear terms become dominant when
 $\delta > \beta^2/H^2 G_X^2 \dot{\phi}^2 \sim \mathcal{O} (1)$. 
Therefore, for $\delta > 1$, 
we need to take into account the non-linear terms, 
and these terms play an important role 
in the Vainshtein mechanism as we will see in section 
\ref{VainshteinMechanism}.

In the limit of $n=\infty$, the linear cosmological perturbation
approaches the $\Lambda$CDM model. 
It would be interesting to clarify what is the nature of the
gravity in the model with large $n$ on the small scales 
where the Vainshtein mechanism works in the static limit. 
However, this is outside of the scope of the present paper, and
we only consider the case $n\simlt10$ in the latter part of
this paper. 

%{\bf
%However, for large $n$, the non-linear term can be neglected
%since it is of order $n^0$. 
%For the case $n=\infty$, the cosmological perturbation exactly matches 
%to the $\Lambda{\rm CDM}$ model, therefore, we do not have to 
%worry about the solar system constraint.
%On the other hand, for large $n$ (not $n=\infty$),
%the Vainshtein mechanism might no longer works,
%but this problem is beyond the scope of this paper.}

%%%%%%%%%%%%%%%%%%%%%%%%%%%%%%%%%%%%%%%%%
\section{Cosmological Constraints from the Redshift-space Distortion}
\label{PS}
%%%%%%%%%%%%%%%%%%%%%%%%%%%%%%%%%%%%%%%%%
\subsection{Current Constraint}
The multipole power spectrum is useful for measuring the redshift-space
distortion, which plays a vital role of testing gravity on the scales
of cosmology \cite{Lindergr,Guzzo,Yamamoto08,WSP,Reyes,Hirano}.
In this section, utilizing the monopole and quadrupole spectra
from the SDSS LRG sample of the data release 7 \cite{devonv,frr},
we investigate a constraint on the kinetic braiding model.
The multipole power spectrum is defined by
\begin{eqnarray}
  P(k,\mu)=\sum_{l=0,2,4,...} P_l(k) \mathcal{L}_l(\mu)(2l+1),
\end{eqnarray}
where $P(k,\mu)$ the redshift-space power spectrum, 
$\mathcal{L}_l(\mu)$ are the Legendre polynomials, 
$\mu$ is the directional cosine between the line of sight direction 
and the wavenumber vector. 
The monopole $P_0(k)$ represents the angular averaged power spectrum, 
and $P_2(k)$ is the quadrupole spectrum which gives the leading 
anisotropic contribution. 
For the theoretical modeling of the redshift-space power spectrum, 
we adopt the fitting formulas developed by Jennings et al. \cite{Jennings}
and Peacock and Dodds \cite{PD}, 
with the transfer function by Eisenstein and Hu \cite{EHu}. 
%For definite, 
We consider the galaxy redshift-space power spectrum 
\cite{Scoccimarro}, 
\begin{eqnarray}
&&P_{\rm gal}(k,\mu)=\left(
b^2(k)P_{\delta\delta}(k)+2fb(k)P_{\delta\theta}(k)\mu^2
+f^2P_{\theta\theta}(k)\mu^4 \right)e^{-(fk\mu\sigma_v)^2},
\label{JE}
\end{eqnarray}
where $P_{\delta\delta}(k)$ is the nonlinear matter power spectrum,
$P_{\theta\theta}(k)$ is the power spectrum of the velocity 
divergence, and $P_{\delta\theta}(k)$ is the cross power 
spectrum of matter and the velocity divergence, 
$b(k)$ is the clustering bias, and $\sigma_v$ is the velocity dispersion.
Jennings et al. proposed a fitting formula for
the redshift-space power spectrum of the form (\ref{JE}), 
assuming $b(k)=1$. The fitting formula relates the nonlinear
matter power spectrum $P_{\delta\delta}(k)$ to $P_{\delta\theta}(k)$ 
and $P_{\theta\theta}(k)$. Although the accuracy of the fitting 
formula for the kinetic braiding modified model is not 
verified, we assume its validity, and use it in this section.
Also we consider the scale dependent bias in the form, 
\begin{eqnarray}
  &&b(k)=b_0+b_1\left(\frac{k}{0.1h{\rm Mpc}^{-1}}\right)^{b_2},
\label{bias}
\end{eqnarray}
where $b_0$, $b_1$, and $b_2$ are the free parameters.
The multipole power spectrum is measured assuming the fiducial
distance-redshift relation $s=s(z)$ of the spatially flat 
$\Lambda$CDM model with $\Omega_0$=0.28.
In order to compare the theoretical prediction of the kinetic 
braiding model with the observed power spectrum, we must take 
the cosmological redshift-space distortion into account \cite{BPH}.
Denoting the comoving distance and redshift relation of the kinetic 
braiding model, $\chi=\chi(z)$, the galaxy power spectrum with the
cosmological redshift-space distortion is 
\begin{eqnarray}
  P_{\rm gal}^{\mathrm{CR}}(k,\mu)=\frac{1}{c_{\parallel} c_{\perp}^2} 
P_{\rm gal}\biggl(
k\rightarrow k\sqrt{{\mu^2\over c_\parallel^2}+{1-\mu^2\over c_\perp^2}},
\mu\rightarrow {\mu/c_\parallel^2\over \sqrt{\mu^2/c_\parallel^2+(1-\mu^2)/c_\perp^2}
}\biggr),
\end{eqnarray}
where we defined $c_{\parallel}=d\chi(z)/ds(z)$, $c_{\perp}=\chi(z)/s(z)$, and
we use the mean redshift $z=0.3$ of the SDSS LRG sample.

We use the monopole and quadrupole power spectra in the range of
wavenumber,  
$0.02h \mathrm{Mpc}^{-1} < k_i <0.2h \mathrm{Mpc}^{-1}$, and compute 
the chi squared,
\begin{eqnarray}
  \hspace{-5mm}\chi^2_{\mathrm{RD}}=\sum_{i,j} \sum_{\ell,\ell'=0,2} (P_\ell(k_i)-P_\ell^{\mathrm{obs}}(k_i)) 
C_{\ell \ell'}^{-1}(k_i,k_j)(P_{\ell'}(k_j)-P_{\ell'}^{\mathrm{obs}}(k_j)),
\label{chi2}
\end{eqnarray}
where $P_\ell^{\mathrm{obs}}(k_i)$ is the observed power spectrum and 
$C_{\ell \ell'}(k_i,k_j)$ is the covariance matrix. 
$P_\ell^{\mathrm{obs}}(k_i)$ and $C_{\ell \ell'}(k_i)$ are from refs.
\cite{devonv}, which used luminous red galaxy sample of 
the Sloan Digital Sky survey data release 7
with the contiguous part of $~7150$ $\mathrm{deg}^2$ sky coverage
in the redshift range $0.16<z<0.47$, and the corresponding 
$1000$ mock catalogues. 

\begin{figure}[]
  \begin{tabular}{rl}
   \begin{minipage}{0.5\textwidth}
    \begin{center}
     \includegraphics[scale=1.4]{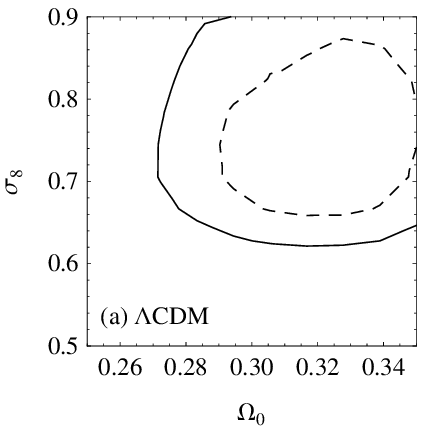}
    \end{center}
   \end{minipage}
   \begin{minipage}{0.5\textwidth}
    \begin{center}
     \includegraphics[scale=1.4]{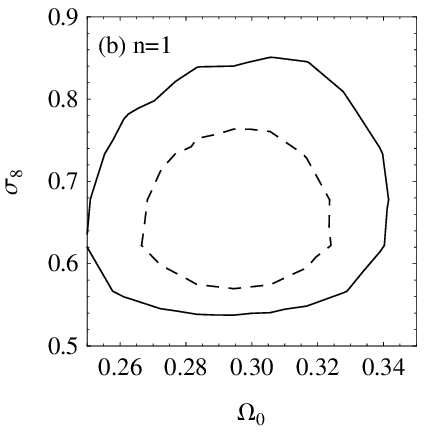}
    \end{center}
   \end{minipage}
\\
   \begin{minipage}{0.5\textwidth}
    \begin{center}
     \includegraphics[scale=1.4]{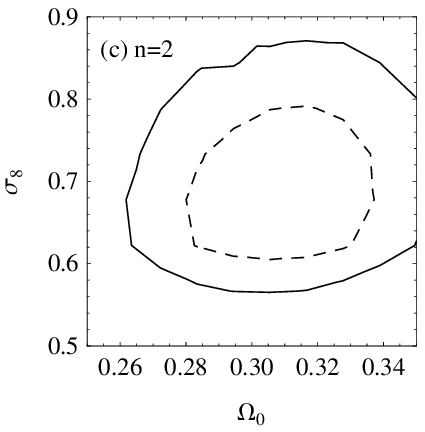}
    \end{center}
   \end{minipage}
   \begin{minipage}{0.5\textwidth}
    \begin{center}
     \includegraphics[scale=1.4]{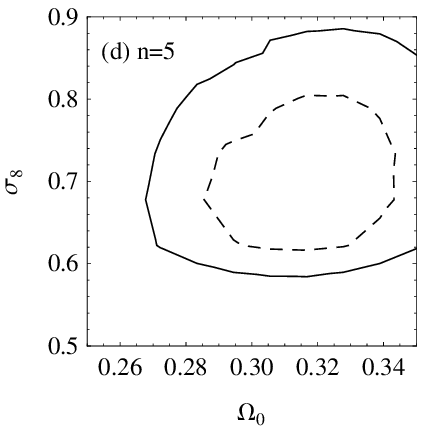}
    \end{center}
   \end{minipage}
  \end{tabular}
  \caption{Contour of $\Delta \chi^2_{\mathrm{RD}}$ on the $\Omega_0 - \sigma_8$ 
plane. The upper left panel (a) is for the $\Lambda \mathrm{CDM}$ model. 
The upper right panel (b) is the kinetic braiding model with $n=1$, 
the lower left panel (c) is $n=2$, and the lower right panel (d) is $n=5$, 
respectively. In each panel, the dashed curve and solid curve are
the $1\sigma$ and $2\sigma$ confidence contour-levels, respectively. 
  \label{psContour}{}}
\end{figure}

Figure~\ref{psContour} shows the contours $\Delta \chi^2_{\mathrm{RD}}$ on the 
$\Omega_0 - \sigma_8$ plane for the $\Lambda \mathrm{CDM}$ model (upper 
left panel) and for the kinetic braiding model with $n=1$ 
(upper right panel), $n=2$ (lower left panel) and $n=5$ (lower right panel),
respectively. In each panel, the $1\sigma$ and $2\sigma$ confidence 
contour-levels 
are given by dashed curve and solid curve, respectively. 
In the analysis, we fixed the spectral index $n_s=0.96$ and the 
density parameter for baryon $\Omega_bh^2=0.0225$, and the $\chi^2_{\mathrm{RD}}$ 
is computed to minimize (\ref{chi2}) by fitting the bias 
parameters $b_0$, $b_1$, $b_2$ and the velocity dispersion 
$\sigma_v$. 
In this figure, we may understand that the constraint on $\Omega_0$ 
comes from the baryon acoustic oscillation feature, and the constraint 
on $\sigma_8$ comes from the amplitude of the monopole and the 
quadrupole spectrum. From figure~\ref{psContour}, the kinetic 
braiding model with smaller $n$ favors the smaller values 
of $\Omega_0$ and $\sigma_8$, compared with 
the $\mathrm{\Lambda CDM}$ model. However, 
the difference is not quite large, and the constraint is not 
very tight.

Our result is obtained by extrapolating the quasi-nonlinear formula 
to the kinetic braiding model. The Vainshtein mechanism 
is important on small scales in the kinetic braiding model, 
as in the case of the DGP model 
\cite{Koyama09,Schmidt09a,ScoccimarroNbody}.
In the present paper, however, we neglect the nonlinear 
Vainshtein effect because we need to consider 
only rather large scales, $k\simlt0.2h{\rm Mpc}^{-1}$.
But, the validity will be necessary to be checked 
using N-body simulations in future.

%%%%%%%%%%%%%%%%%%%%%%%%%%
\subsection{Future Prospect}
\label{mpsfp}
%%%%%%%%%%%%%%%%%%%%%%%%%%
We now consider a future prospect constraining the kinetic 
braiding model using the multipole power spectra. Let us 
consider a redshift survey like the SuMIRe (SUbaru Measurement 
of Imaging and REdshift of the universe) PFS 
(Prime Focus Spectrograph) survey \cite{sumire}, 
which assumes the survey parameters like those of the WFMOS 
survey \cite{WFMOS}. 
In the SuMIRe PFS survey, we assume the sky coverage is 
$2000$ $\mathrm{deg}^2$ 
over redshift range $0.8<z<1.6$ with the averaged number density of 
galaxy  $\bar{n}=3 \times 10^{-4}h^3\mathrm{Mpc}^{-3}$. 

\begin{figure}[t]
  \begin{center}
    \includegraphics[scale=1.6]{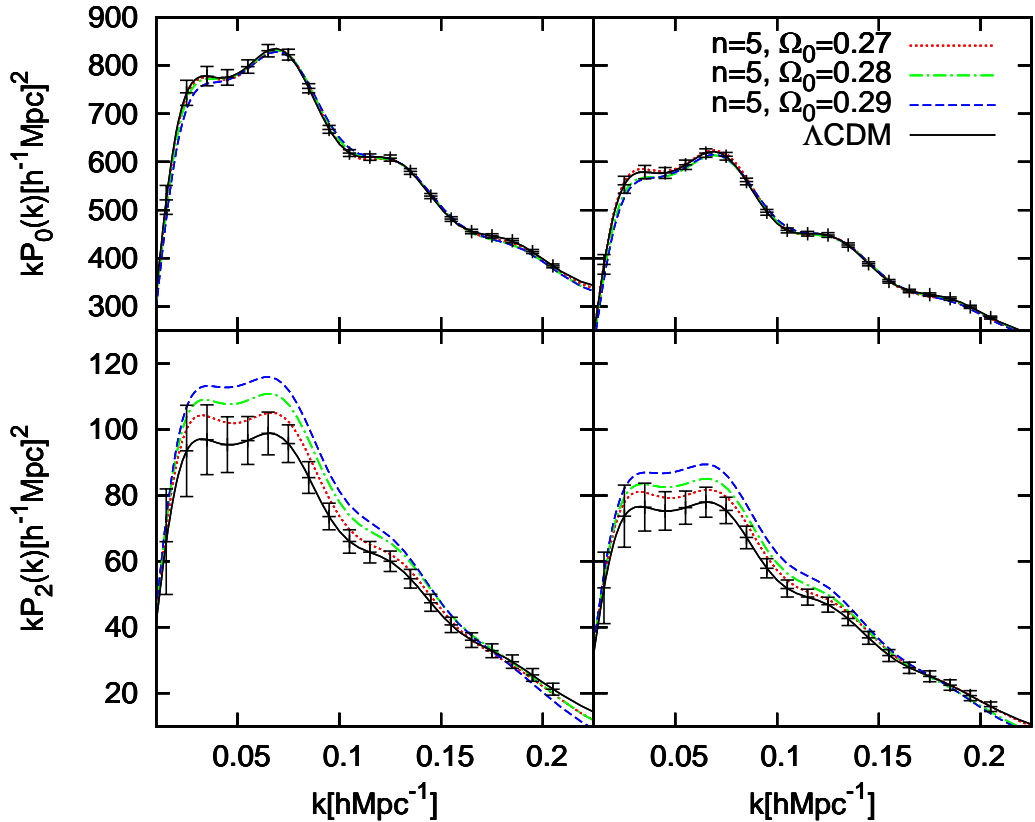}% 
  \end{center}
  \caption{ Monopole (upper panels) and quadrupole (lower panels) power 
spectra multiplied by $k$, $kP_0(k)$ and $kP_2(k)$ for the 
fiducial $\Lambda \mathrm{CDM}$ model and for the kinetic braiding model 
of $n=5$ 
with $\Omega_0=0.27,~0.28$, and $0.29$. The left panels are the subsample 
of the range of redshift $0.8<z<1.2$, while the right panels are 
the subsample of the range redshift $1.2<z<1.6$. 
The error bars is estimated using eq.~(\ref{error}). 
\label{sumire} {}}
\end{figure}

Figure~\ref{sumire} shows the monopole (upper panels) and 
quadrupole (lower panels) power spectra multiplied by $k$, 
$kP_0(k)$ and $kP_2(k)$ for the fiducial $\Lambda \mathrm{CDM}$ model 
with $\Omega_0h^2=0.1344$, and for the kinetic braiding model with 
$\Omega_0=0.27,~0.28$, and $0.29$.
Here, we fixed $n=5$. 
The left panels assume the subsample of the 
the range of redshift $0.8<z<1.2$, while the right panels 
assume the subsample of the range redshift $1.2<z<1.6$. 
The theoretical modeling of the power spectrum is the same as that 
in the above subsection, which is evaluated at the mean redshift of
each subsample.
In computing the theoretical power spectrum, we 
fixed the initial amplitude of the fluctuation so that the matter 
power spectrum of the $\rm{\Lambda CDM}$ model gives $\sigma_8=0.8$.
Other cosmological parameters are fixed as $n_s=0.96$ and 
$\Omega_bh^2=0.0225$.
The cosmological redshift-space distortion is taken in the 
power spectrum of the kinetic braiding model into account, 
assuming that the distance-redshift relation of the 
$\rm{\Lambda CDM}$ model is adopted in data analysis. 
The error bars for the $\rm{\Lambda CDM}$ model is evaluated by \cite{Nakamichi}
\begin{eqnarray}
  \Delta P_\ell(k)^2=\frac{(2\pi)^3}{\Delta V_k} \int^{1}_{-1} d\mu 
\frac{\left[\mathcal{L}_l(\mu)\right]^2}
  {V \bar{n}^2 \left[1+\bar{n}
P(k,\mu) \right]^{-2}},
\label{error}
\end{eqnarray}
where $V_k$ is the volume of a shell in the Fourier space, 
and $V$ is the survey volume, $V={\cal A} \int_{z_{\rm min}}^{z_{\rm max}} 
dz (ds/dz) s^2$, where ${\cal A}$ is the survey area, and 
$z_{\rm max}$ and $z_{\rm min}$ are
the maximum and minimum redshifts of the survey, respectively.

As for the bias $b(k)$, We adopted the scale-independent bias $b=1.5$ 
and the  velocity dispersion $\sigma_v=350 \mathrm{km/s}$ for 
the $\Lambda \mathrm{CDM}$ model, but the scale-dependent
bias (\ref{bias}) for the kinetic braiding model. 
In computing the power spectrum of kinetic braiding model, we 
assumed $b_0,~b_1,~b_2$ and $\sigma_v$ as free parameters
which are determined so as to minimize the value 
\begin{eqnarray}
\sum_i\sum_{\ell=0,2}[P_\ell^{\rm \Lambda CDM}(k_i)-P_\ell^{\rm Braiding}(k_i)]^2/\Delta P_\ell(k_i)^2,
\end{eqnarray}
where $P_\ell^{\rm \Lambda CDM}(k_i)$ and $P_\ell^{\rm Braiding}(k_i)$ are the 
power spectrum of the $\Lambda$CDM model and the kinetic braiding model, 
and we used $0.02h \mathrm{Mpc}^{-1}< k_i<0.2h \mathrm{Mpc}^{-1}$.

Figure~\ref{sumire} shows that the monopole spectrum can be
almost same in the different models because of the bias,
but the quadrupole spectrum is different. 
Thus, the combination of the monopole 
and quadrupole power spectra is useful to distinguish between 
the $\Lambda$CDM model and the kinetic braiding model, depending 
on  the model parameters. In \ref{FMAPK}, we estimates a constraint 
on $\Omega_0$ and $n$ at a future quantitative level with the Fisher matrix 
analysis. 

% Fig.~\ref{sumire} shows that it might be difficult to distinguish
% between the $\Lambda$CDM model and the kinetic braiding model only
% by a measurement of the monopole spectrum,
% but there can be a chance to do it by combining with the quadrupole
% power spectrum in the future survey.

%%%%%%%%%%%%%%%%%%%%%%%%%%%%%%%%%%%%%%%%%%%%%%%%%%%%%%%%%%
\subsection{Fisher Matrix Analysis}
\label{FMAPK}
\begin{figure}[t]
  \begin{center}
    \includegraphics[scale=1.3]{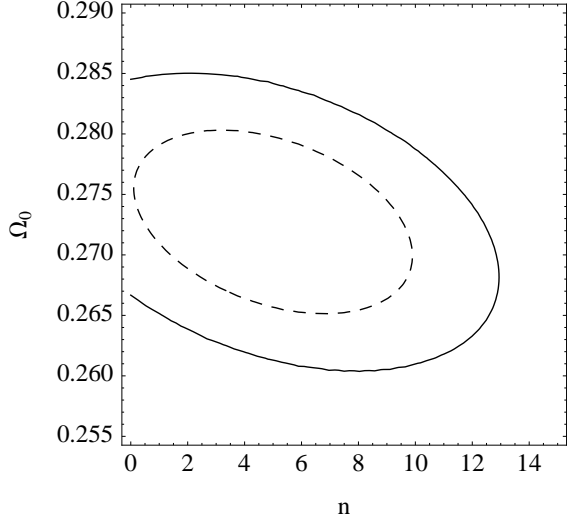}% 
  \end{center}
  \caption{
The 1-sigma (dashed curve) and 2-sigma (solid curve) contours in 
the $\Omega_0 - n$ plane when we assumed the WFMOS/SuMIRe PFS 
survey, ${\cal A}=2000$ $\mathrm{deg}^2$, 
the range of the redshift $0.8\leq z\leq 1.6$ and the averaged number 
density of galaxy $\bar{n}=3 \times 10^{-4}h^3\mathrm{Mpc}^{-3}$. 
%The contours are obtained by marginalizing the 
%Fisher matrix for the galaxy power spectrum with the 6 parameters, 
%$n$, $\Omega_0$, $b_0$, $b_1$, $b_2$ and $\sigma_v$ over other 
%than $n$ and $\Omega_0$.
The target model is the kinetic braiding model with 
$n=5$, $\Omega_0h^2=0.1344$, %$h=0.7$, 
$b_0=2$, $b_1=0.1$, $b_2=0.1$, and $\sigma_v=350 {\rm km/s}$.
We also fixed $\Omega_bh^2=0.0225$, $n_s=0.96$ and  
the amplitude of the initial cosmological perturbation 
so that $\sigma_8=0.8$ for the $\Lambda$CDM model.
}
\label{fig_fisher_ps} {}
\end{figure}

%%%%%%%%%%%%%%%%%%%%%%%%%%%%%%%%%%%%%%%%%%%%%%%%%%%%%%%%%%
In this subsection, let us discuss the future prospect of a constraint 
with galaxy power spectrum quantitatively using Fisher matrix analysis.  
We adopt the following Fisher matrix, for simplicity, 
\begin{eqnarray}
F_{ij}&=&{1\over 4\pi^2}\int_{k_{\rm min}}^{k_{\rm max}} dk k^2 \int_{-1}^{+1}d\mu
{\partial \bigl<P^{\rm CR}_{\rm gal}(k,\mu)\bigr>\over \partial \theta_i} 
{\partial \bigl<P^{\rm CR}_{\rm gal}(k,\mu)\bigr>\over \partial \theta_j} 
%\nonumber
%\\
%&&~~\times
{V\over (P^{\rm CR}_{\rm gal}(k,\mu)+1/\bar n)^2},
\nonumber\\
\end{eqnarray}
where $V$ is a survey volume, ${\bar n}$ is a mean number density of 
galaxies, $\theta_i$ denotes a model parameter.
In this Fisher matrix analysis, we consider the 6 parameters 
$n$, $\Omega_0$, $b_0$, $b_1$, $b_2$ and $\sigma_v$.
The amplitude of the initial cosmological perturbation is fixed
so that $\sigma_8=0.8$ for the $\Lambda$CDM model.
We also fixed the spectral index $n_s=0.96$ and $\Omega_bh^2=0.0225$.  
The curves in figure~\ref{fig_fisher_ps} show the 1-sigma (dashed 
curve) and 2-sigma (solid curve) confidence contours in the $n$ and 
$\Omega_0$ plane, by marginalizing the Fisher matrix of the 6 parameters, 
$n$, $\Omega_0$, $b_0$, $b_1$, $b_2$ and $\sigma_v$ over
the parameters other than  $n$ and $\Omega_0$.
In figure~\ref{fig_fisher_ps}, we assumed the redshift survey like 
WFMOS/SuMIRe PFS survey, adopted in the previous subsection. 
The sky coverage is $2000$ $\mathrm{deg}^2$, the minimum and the 
maximum redshifts are $z_{\rm min}=0.8$ and $z_{\rm max}=1.6$, respectively, 
and the averaged number density of galaxy  is 
$\bar{n}=3 \times 10^{-4}h^3\mathrm{Mpc}^{-3}$. 
The target model is the kinetic braiding model with 
$n=5$, $\Omega_0h^2=0.1344$, $b_0=2$, $b_1=0.1$, $b_2=0.1$, 
and $\sigma_v=350 {\rm km/s}$. %$h=0.7$, 
We assumed that the power spectrum is measured adopting the fiducial
distance-redshift relation $s=s(z)$ of the spatially flat 
$\Lambda$CDM model with $\Omega_0$=0.28.%, and $\sigma_8=0.8$.

Figure~\ref{fig_fisher_ps} implies that the constraint on $\Omega_0$ 
is tight, while the constraint on $n$ is not very tight.
However, the 1-sigma (2-sigma) error in determining $n$ is
$\Delta n\sim 5$ ($\Delta n\sim10$) when the target parameter 
is $n=5$. Then, it is possible to distinguish between the
kinetic braiding model and the $\Lambda$CDM model if $n\simlt 10$.
However, the constraint on $n$ becomes weaker when the target 
value of $n$ becomes large,
since the background expansion and cosmological perturbations 
in the kinetic braiding model for $n\simgt1000$ is almost 
the same as the $\Lambda{\rm CDM}$ model.
Then, it will be difficult to distinguish between the $\Lambda$CDM 
model and the kinetic braiding model by a measurement of the 
redshift-space power spectrum when $n$ is large.

\section{Non-linear Density Perturbation}
\label{VainshteinMechanism}
In this section, we investigate the nonlinear effect of the perturbed galileon 
field on small scales. 
%{\bf
%Since the quasi-approximation is valid for small $n$, 
%We consider the nonlinear effect for $n<10$.
%}
As discussed at the end of section 
\ref{LinearPerturbation}, the non-linear term 
in the perturbed galileon field equation (\ref{NonlinearFieldEq}) becomes 
important for $\delta\simgt1$.
Therefore, when considering the non-linear cosmological perturbations, we need 
to solve eq.~(\ref{NonlinearFieldEq}). However, it is not easy to solve 
eq.~(\ref{NonlinearFieldEq}) in general.
For simplicity, we assume that the density perturbation 
$\delta \rho$ has a spherically symmetric distribution and a top-hat profile. 
Then, the perturbation equation for the galileon field can be solved analytically, 
and it gives a qualitative picture of the Vainshtein mechanism in the kinetic 
braiding model. 
Note that eq.~(\ref{NonlinearFieldEq}) is identical to the equation for the 
brane bending mode $\varphi$ in the DGP model \cite{Koyama07,Lue04,Schmidt09a} 
and the full profile of the brane bending mode $\varphi$ is investigated 
in \cite{Schmidt10}. 
The behavior of $\varphi$ in a top-hat density profile remains the same 
in the kinetic braiding model.

We start with recasting the Poisson equation (\ref{Poisson}) and the field equation (\ref{NonlinearFieldEq}) by replacing the comoving 
nabla operator with the physical nabla operator, then we have
\begin{eqnarray}
  &&\nabla^2\varphi+\lambda^2(a)\left[(\nabla_i\nabla_j\varphi)(\nabla^i\nabla^j\varphi)
-(\nabla^2\varphi)^2\right]=-4 \pi G \zeta(a) \delta \rho,
\label{NonlinearFieldEq2}
  \\
  &&\nabla^2\Psi=4\pi G \rho \delta- \xi(a) \nabla^2\varphi,
\label{Poisson2}
\end{eqnarray}
where we defined
\begin{eqnarray}
  &&\lambda^2(a)=\frac{G_X\phi}{\beta(a)}, \hspace{1cm}
%\\
%  &&
\zeta(a)=\frac{G_X\dot{\phi}^2}{\beta(a)\phi}, \hspace{1cm}
%\\
%  &&
\xi(a)=4\pi G G_X\dot\phi^2\phi.
\end{eqnarray}
As we will show below, the following combinations are useful, 
\begin{eqnarray}
&&\xi(a)\zeta(a)={(1-\Omega_{\rm m}(a))(2n-\Omega_{\rm m}(a))\over 
\Omega_{\rm m}(a)(5n-\Omega_{\rm m}(a))},
\\
&&\lambda^2(a)\zeta(a)=\left({2n-\Omega_{\rm m}(a)\over
H(a)\Omega_{\rm m}(a)(5n-\Omega_{\rm m}(a))}\right)^2.
\end{eqnarray}
Under the assumption that the density perturbation $\delta \rho (r)$ is spherically symmetric, 
the field equation (\ref{NonlinearFieldEq2}) can be written as
\begin{eqnarray}
  \frac{1}{r^2}\frac{d}{dr}\left[r^2\frac{d\varphi}{dr}\right]-\frac{2\lambda^2(a)}{r^2}\frac{d}{dr}\left[r\left(\frac{d\varphi}{dr}\right)^2\right]=-4 \pi G \zeta(a) \delta\rho.
\label{NonlinearFieldEq3}
\end{eqnarray}
Then, we define the enclosed mass perturbation,
$  m(r)=4\pi \int^r_0 r'^2\delta\rho(r')dr'$.
Integrating eq.~(\ref{NonlinearFieldEq3}) over $r^2dr$, we have
\begin{eqnarray}
  r^2\frac{d\varphi}{dr}-2\lambda^2(a) r\left(\frac{d\varphi}{dr}\right)^2=-\zeta(a) Gm(r).
\label{dphidr}
\end{eqnarray}
Solving for $d\varphi/dr$, we have
\begin{eqnarray}
  \frac{d\varphi}{dr}=-\frac{r_V}{4\lambda^2(a)}g\left(\frac{r}{r_V}\right),
\end{eqnarray}
where $g(x)=x\left[\sqrt{1+x^{-3}}-1\right]$ and the Vainshtein radius of the enclosed mass is 
\begin{eqnarray}
  r_V=\left[8\lambda^2(a)\zeta(a) Gm(r)\right]^{1/3} =\left[\frac{8G_X^2\dot{\phi}^2 Gm(r)}{\beta^2(a)} \right]^{1/3}.
  \label{VainshteinRadius}
\end{eqnarray}
Practically, eq.~(\ref{dphidr}) yields two branches 
of solutions, which correspond to the attractive and 
repulsive force. We require that the solution matches 
the linear solution, given by 
$d\varphi/dr \propto r^{-2}$ at $r \gg r_V$, hence 
we pick the solution whose value vanishes at an infinite distance.
The Vainshtein radius for a point source can be estimated using 
eq.~(\ref{VainshteinRadius}) and (\ref{Att}), 
\begin{eqnarray}
  r_V&=&\left(\frac{4r_g}{9 H^2 \beta^2}\right)^{1/3} 
%  r_V=\left(\frac{4 \alpha^2 r_g }{9 H^2 \beta^2}\right)^{1/3} 
\nonumber
\\
     &=&1.91 \times 10^2 \left[\frac{2n-\Omega_m(a)}{\Omega_m(a)(5n-\Omega_m(a))}\right]^{2/3} 
      \left(\frac{H}{H_0}\right)^{-2/3} \left(\frac{M}{M_{\odot}} \right)^{1/3} \mathrm{pc},
\end{eqnarray}
where $r_g$ is the Schwartzschild radius of the point mass. 
Using the present values of $H$ and $\Omega_m$ for $n=1$, 
the Vainshtein radius is approximated by $r_V=2.3 \times 10^2 (M/M_{\odot})^{1/3} \mathrm{pc}$. 
The asymptotic behavior of the galileon field well inside and outside the Vainshtein radius is 
\begin{eqnarray}
  \frac{d\varphi}{dr}= \left\{ \begin{array}{ll}
    -\zeta(a) \displaystyle{\frac{Gm}{r^2}} & r \gg r_V, \\
    \\
    -\displaystyle{\frac{r_V^{3/2}}{4 \lambda^2(a) r^{1/2}}}& r \ll r_V, \\
\end{array} \right.
\end{eqnarray}
respectively.
Well outside the Vainshtein radius, $r \gg r_V$, $d\varphi/dr$ 
is proportional to the Newtonian potential 
and this solution corresponds to the linear solution. 
On the other hand, well inside the Vainshtein radius, 
$r \ll r_V$, the galileon field $\varphi$ approaches 
nearly constant \cite{Schmidt10} and does not contribute 
to the modified Poisson equation (\ref{Poisson2}).

%%%%%%%%%%%%%%%%%%%%%%%%%
\subsection{Top-Hat Profile}
%%%%%%%%%%%%%%%%%%%%%%%%%
Hereafter, we consider the following top-hat density profile $\rho(r)$ or, equivalently,
 the mass perturbation $m(r)$
\begin{eqnarray}
  \rho(r)= \left\{ \begin{array}{ll}
    \bar{\rho}+\delta\rho & ~~~~r\leq R \\
    \bar{\rho} & ~~~~r>R\\
\end{array} \right.,
%\end{eqnarray}
%\begin{eqnarray}
\hspace{1cm}
  m(r)= \left\{ \begin{array}{ll}
    \delta M (r/R)^3 & ~~~~r\leq R \\
    \delta M & ~~~~r>R\\
\end{array} \right..
\end{eqnarray}
We neglect any compensation underdensity swept out by the prior evolution 
of the top hat, and assume that a density enhancement is constant.
For $r \leq R$, $d\varphi/dr$ is proportional to $r$. 
Then, the non-linear term in eq.~(\ref{NonlinearFieldEq2}) can be written as
\begin{eqnarray}
  (\nabla_i\nabla_j\varphi)(\nabla^i\nabla^j\varphi)-(\nabla^2\varphi)^2=-\frac{2}{3}(\nabla^2\varphi)^2.
\end{eqnarray}
Therefore, eq.~(\ref{NonlinearFieldEq2}) can be written as 
\begin{eqnarray}
  \nabla^2\varphi-\frac{2\lambda^2(a)}{3}(\nabla^2\varphi)^2=-4 \pi G \zeta(a) \delta\rho.
\end{eqnarray}
Setting $r=R$ and solving for $\nabla^2\varphi$, we have 
\begin{eqnarray}
  \nabla^2\varphi=-8\pi G \zeta(a) \left(\frac{R}{R_V}\right)^2 g\left(\frac{R}{R_V}\right)\delta\rho,
\end{eqnarray}
where $R_V=\left[8\lambda^2(a)\zeta(a) G \delta M\right]^{1/3}$.
Then, the modified Poisson equation (\ref{Poisson}) becomes
\begin{eqnarray}
  \nabla^2\Psi=4\pi G_{\mathrm{eff}}^{\mathrm{NL}}\rho\delta,
\end{eqnarray}
where the non-linear effective gravitational constant $G_{\mathrm{eff}}^{\mathrm{NL}}$ is given by
\begin{eqnarray}
  4\pi G_{\mathrm{eff}}^{\mathrm{NL}}=4\pi G\left[1+2\xi(a)\zeta(a)\left(\frac{R}{R_V}\right)^2 
g\left(\frac{R}{R_V}\right) \right].
\label{GeffNonlinear}
\end{eqnarray}
The first term of the right-hand-side of eq.~(\ref{GeffNonlinear}) is the contribution from 
the Newtonian gravity and the second term induces the additional 
attractive force due to 
the perturbed galileon field. 
In the case $R \to 0$, corresponding to $\delta \rho \to \infty$, 
the effective gravitational constant $G_{\mathrm{eff}}^{\mathrm{NL}}$ 
approaches the 
standard gravitational constant $G$ and this is the Vainshtein mechanism in 
the spherical collapse in the cosmological perturbations 
(See the solid curve in the right panel of figure~\ref{SCollaseFig}). 

%%%%%%%%%%%%%%%%%%%%%%%%%%
\subsection{Spherical Collapse}
%%%%%%%%%%%%%%%%%%%%%%%%%%
In this section, we examine the spherical collapse which is a powerful tool for 
understanding the formation of a bound system in non-linear regime 
\cite{Gunn,Lahav91,Wang98,Horellou05,Basilakos09,mota06}. 
The evolution of a spherical collapse undergoes three phases: (i) turn around, 
(ii) collapse, and (iii) virialization. 
A spherical overdense region expands along the expansion of the universe 
up to the maximum radius as the density perturbation evolves proportionally
to the scale factor at early times. 
Then, it starts to shrink toward the center of the spherical region and 
finally collapses. The spherical collapse approach enables us to evaluate 
the critical density contrast $\delta_c$ and the virial density 
$\Delta_{\mathrm{vir}}$, which need in predicting the mass function of the halos. 
We assume that perturbed matter distribution remains a top-hat profile and 
each shell does not cross during collapse. 

We start with the fully non-linear evolution equation for the matter over-density, 
which can be obtained from the energy-momentum conservation,
\begin{eqnarray}
  \ddot{\delta}-\frac{4}{3}\frac{\dot{\delta}^2}{1+\delta}+2H\dot{\delta}=(1+\delta)\nabla^2\Psi.
\label{MatterEvolution}
\end{eqnarray}
We assume that the total mass inside $R$ is conserved during collapse, 
\begin{eqnarray}
  M=\frac{4\pi}{3}R^3\rho=\frac{4\pi}{3}R^3\bar{\rho}(1+\delta)=\mathrm{constant}.
\label{MassConservation}
\end{eqnarray}
Differentiating eq.~(\ref{MassConservation}) with respect to time gives
\begin{eqnarray}
  \frac{\ddot{R}}{R}&=&H^2+\dot{H}-\frac{1}{3(1+\delta)}\left[\ddot{\delta}+2H\dot{\delta}-\frac{4}{3}\frac{\dot{\delta}^2}{1+\delta}\right].
\end{eqnarray}
Combining (\ref{MatterEvolution}) and (\ref{MassConservation}), 
the evolution equation for the spherical collapse is given by
\begin{eqnarray}
  \frac{\ddot{R}}{R}&=&H^2+\dot{H}-\frac{1}{3}\nabla^2\Psi \nonumber\\
  &=&H^2+\dot{H}-\frac{4\pi G_{\mathrm{eff}}^{\mathrm{NL}}}{3}\rho\delta,
\label{SphericalCollapseEq}
\end{eqnarray}
where we used (\ref{GeffNonlinear}) in the second line. 
For convenience, we rewrite the equation of motion as 
\begin{eqnarray}
  \frac{\ddot{R}}{R}=-\frac{1}{3}\left[\nabla^2\Psi_{\mathrm{b}}+\nabla^2\Psi_{\mathrm{g}}\right],
\end{eqnarray}
where we have defined the gravitational potential contributed from 
the effective background term $\nabla^2\Psi_{\mathrm{b}}=-3(H^2+\dot{H})$ 
and from the effective gravitational constant 
$\nabla^2\Psi_{\mathrm{g}}=4 \pi G_{\mathrm{eff}}^{\mathrm{NL}}\rho\delta$.

%---------------------------------------------------------
\subsection{Virial Theorem}
%---------------------------------------------------------
For a spherically symmetric top-hat model, the total potential energy from the effective background term 
and the gravity is given by
\begin{eqnarray}
  W=\sum_{s=\mathrm{b},\mathrm{g}} W_s=-3M\int^R_0 \frac{r^2 dr}{R^3}r \sum_{s=\mathrm{b},\mathrm{g}}\frac{d\Psi_s}{dr},
  \label{PotentialEnergyS}
\end{eqnarray}
where the potential energy of the gravity and the effective background term
are 
\begin{eqnarray}
  &&W_{\mathrm{g}}=-\frac{3}{5}\frac{G_{\mathrm{eff}}^{\mathrm{NL}} M \delta M}{R},
  \label{PotentialEnergyN}\\
  &&W_{\mathrm{b}}=\frac{3}{5}(H^2+\dot{H})MR^2,
  \label{PotentialEnergyEFF}
\end{eqnarray}
respectively. The virial theorem can be obtained by integrating the Boltzmann 
equation over all position, which gives $2T_{\mathrm{vir}}+W_{\mathrm{vir}}=0$.
The virial radius $R_{\rm{vir}}$ is usually calculated by requiring 
the energy conservation at turnaround and virialization. 
However, there is the energy non-conservation problem in the dark
 energy models, DGP and other modified gravity models \cite{Schmidt10, Maor05}.
The violation of energy conservation also arises 
in the kinetic braiding model due to the time-varying 
$\rho_{\phi}$ and $G_{\rm{eff}}$.
Hence, we need carefully choose the condition for the virialization 
epoch. 
In the present paper, we follow the approach in \cite{Schmidt10}.
We choose the virialization epoch so that the virial condition 
satisfied, $2T_{\rm vir}(a_{\rm vir})+W(a_{\rm vir})=0$, 
where the kinetic energy during collapse can be computed by
\begin{eqnarray}
  T=\int d^3x \frac{1}{2}\rho v^2 = \frac{3}{10}M \dot{R}^2.
\end{eqnarray}
We define the virial radius $R_{\mathrm{vir}}$ as the radius 
at the virialization epoch. 
The virial density is given by
\begin{eqnarray}
  \Delta_{\mathrm{vir}}=\frac{\rho_{\rm{vir}}}{\rho_{\rm{collapse}}}
  =[1+\delta(R_{\mathrm{vir}})]
  \left(\frac{a_{\mathrm{collapse}}}{a_{\mathrm{vir}}}\right)^3,
\end{eqnarray}
where $\rho_{\rm{vir}}=\bar\rho(1+\delta)\big|_{a=a_{\rm vir}}$ 
is the density at the virialization epoch 
and $\rho_{\rm{collapse}}$ is the background matter density 
at the time of collapse.

\begin{figure}[t]
  \begin{tabular}{cc}
   \begin{minipage}{0.5\textwidth}
    \begin{center}
     \includegraphics[scale=1.4]{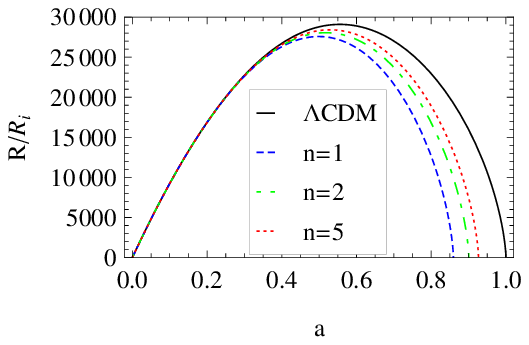}
    \end{center}
   \end{minipage}
   \begin{minipage}{0.5\textwidth}
    \begin{center}
     \includegraphics[scale=1.4]{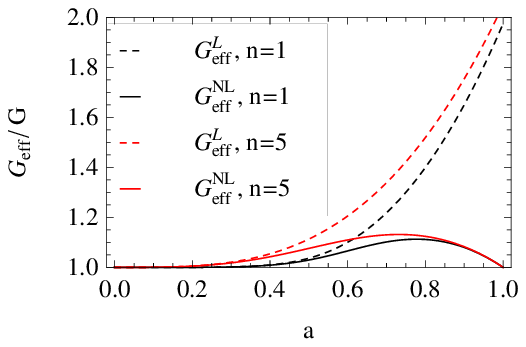}
    \end{center}
   \end{minipage}
  \end{tabular}
  \caption{ Left panel: Evolution of the scaled radius 
$R/R_{\rm i}$ of perturbations as a function of the 
scale factor for $\mathrm{\Lambda CDM}$ (solid curve) 
and the kinetic braiding model with $n=1$ (dashed curve),
$n=2$ (dash-dotted curve), and $n=5$ (dotted curve), respectively. 
Right panel: Evolution of the effective gravitational constant 
$G_{\mathrm{eff}}$ in the linear theory (dashed curve) of 
(\ref{geffomega}) and the 
non-linear theory (solid curve) of (\ref{GeffNonlinear}), 
normalized by $G$, as a function of the scale factor for 
%$\mathrm{\Lambda CDM}$ and this 
the kinetic braiding model with $n=1$ (dark black curve) and 
$n=5$ (light red curves).
  \label{SCollaseFig}{}}
\end{figure}

\begin{figure}[h]
  \begin{tabular}{cc}
   \begin{minipage}{0.5\textwidth}
    \begin{center}
     \includegraphics[scale=1.4]{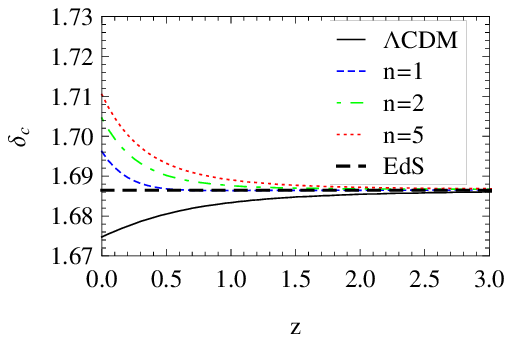}
    \end{center}
   \end{minipage}
   \begin{minipage}{0.5\textwidth}
    \begin{center}
     \includegraphics[scale=1.4]{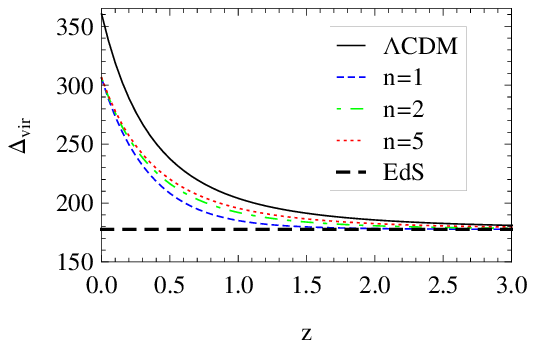}
    \end{center}
   \end{minipage}
  \end{tabular}
  \caption{Left panel: The redshift evolution of the linear overdensity 
for $\mathrm{\Lambda CDM}$ model (solid curve), Einstein de Sitter 
universe (thick-dashed line), and the kinetic braiding model 
as a function of redshift for $n=1$ (dashed curve), 
$n=2$ (dash-dotted curve), and $n=5$ (dotted curve).
Right panel: The redshift evolution of the virial density.
  \label{Deltac} {}}
\end{figure}

We are now ready to solve the spherical collapse 
eq.~(\ref{SphericalCollapseEq}), 
numerically. The total mass conservation within $R$ gives 
\begin{eqnarray}
  \delta(R,a)=\left(\frac{a}{a_{\rm i}}\right)^3
\left(\frac{R}{R_{\rm i}}\right)^{-3}(1+\delta_{\rm i})-1,
\end{eqnarray}
where $R_{\rm i}$ is the initial radius of the perturbation, 
$\delta_{\rm i}$ is the initial density fluctuation, 
and $a_{\rm i}$ is the initial scale factor. 
We set the initial scale factor $a_{\rm i}=10^{-5}$. 
Turnaround and collapse are, respectively, defined by $\dot{R}=0$ and $R=0$. 
The left panel of figure~\ref{SCollaseFig} shows the evolution of the scaled 
radius for the $\Lambda \mathrm{CDM}$ and the kinetic braiding model with 
$n=1,~2$ and $5$. We set the initial density perturbation $\delta_{\rm i}$ 
for the $\Lambda\mathrm{CDM}$ so that collapse occurs at $a=1$. 
For the same initial density perturbation $\delta_{\rm i}$, 
the spherical overdensity region collapses earlier in the kinetic 
braiding model compared with the $\Lambda \mathrm{CDM}$ model 
due to the additional attractive force. 
The right panel of figure~\ref{SCollaseFig} shows the linear and 
non-linear effective gravitational constant as a function of redshift. 
The radius of the overdensity region enters the Vainshtein radius around 
$a \sim 0.48$ for the case $n=1$, and $G_{\mathrm{eff}}^{\mathrm{NL}}$ approaches 
the standard gravitational constant $G$ in the kinetic braiding model. 
Thus, the non-linear effect restores the general relativity
in high density regions through the Vainshtein mechanism. 
Figure~\ref{Deltac} shows the evolution of the critical 
density contrast $\delta_c$ and the virial density
as a function of redshift. The critical density contrast is 
the value of the linear-theory density contrast 
reached to infinite density when the nonlinear 
top-hat perturbation collapses. One can see that $\delta_c$
for the kinetic braiding model approaches that of the Einstein 
de Sitter universe, $\delta_c=3(12\pi)^{2/3}/20$, 
at high redshift. 
The virial density for the kinetic braiding model also approaches 
the value for Einstein de Sitter universe, $\Delta_{\mathrm{vir}}=18\pi^2$, 
at high redshift.
This is because the evolution of background and cosmological perturbations 
in the kinetic braiding model is identical to that in Einstein de Sitter 
universe at early times.
Table \ref{table2} shows the numerical results of 
the critical density contrast $\delta_c$ 
and the virial density $\Delta_{\rm vir}$ at $z=0$ 
for the $\rm{\Lambda CDM}$ model
and the kinetic braiding model with $n=1,~2,~3,~4,$ and $5$.

%%%%%%%%%%%%%%%%%%%%%%%%%%%%%%%%%%%%%%%%%%%%%%%%%%%%%%%%%%%%%%%%%%%%%%%%
\begin{table}[htbp]
 \begin{center}
  \begin{tabular}{l|llllll}
   \hline
   ~~~~~~ &~ ~$\mathrm{\Lambda CDM}$~~~~  & ~ ~$n=1$~~~~  & ~$n=2$~~~~ & ~$n=3$~~~~ & ~$n=4$~~~~ & ~$n=5$
~~~~\\
   \hline
   ~~$\delta_c$        &~ 1.675  &~ 1.696  & 1.705   &  1.708    &  1.710    &  1.711   \\
   ~~$\Delta_{\rm vir}$ &~ 360.4  &~ 305.8  &  306.2  &   306.3   &  306.3    & 306.4   \\
   \hline
  \end{tabular}
  \caption{
    $\delta_c$ and $\Delta_{\rm vir}$ for the kinetic braiding 
    model with $n=1,~2,~3,~4,$ and $5$, respectively, at $z=0$.
\label{table2}}
 \end{center}
\end{table}
%%%%%%%%%%%%%%%%%%%%%%%%%%%%%%%%%%%%%%%%%%%%%%%%%%%%%%%%%%%%%%%%%%%%%%%%

%%%%%%%%%%%%%%%%%%%%%%%%%%%%%%%%%%
\subsection{Galaxy Cluster Counts}
\label{GCC}
%%%%%%%%%%%%%%%%%%%%%%%%%%%%%%%%%%
We here give a simple prediction about the mass function of halos 
by using the Press-Schechter formalism \cite{PressSchechter}.
The number density of virialized clusters in an unit mass range 
as a function of redshift is given by
\begin{eqnarray}
  &&\frac{dn(M,z)}{dM}=\sqrt{\frac{2}{\pi}}\frac{\rho}{M^2}
\left[-\frac{\delta_c(z)}{\sigma(M,z)}
    \frac{d \ln \sigma(M,z)}{d \ln M}\right] 
\exp\left[-\frac{\delta_c^2(z)}{2\sigma^2(M,z)}\right],
\end{eqnarray}
where $M=4 \pi R^3 \rho /3$, and $\sigma(M,z)$ is 
\begin{eqnarray}
  \sigma^2(M,z)=\frac{1}{2\pi^2} \int^\infty_0 k^2 dk P(k,z) W(kR) ^2
\end{eqnarray}
with the top-hat smoothing function $W(kR)=3(\sin kR-kR\cos kR)/(kR)^3$
and the linear matter power spectrum $P(k,z)$ at the redshift $z$.
The number count of halos in unit redshift within the range of 
mass, $M_1 \le M \le M_2$, is given by integrating the halo mass function,
\begin{eqnarray}
  \frac{dN}{dz}=\frac{dV}{dz}\int^{M_2}_{M_1} \frac{dn(M,z)}{dM} dM,
\end{eqnarray}
where $dV/dz= {\cal A}\chi^2(z) d\chi/dz$ is the comoving volume 
of the unit redshift and ${\cal A}$ is the survey area. 

\begin{figure}[t]
  \begin{tabular}{cc}
   \begin{minipage}{0.5\textwidth}
    \begin{center}
     \includegraphics[scale=1.2]{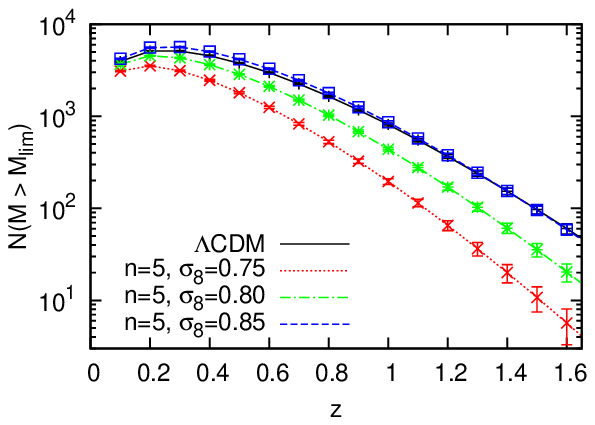}
    \end{center}
   \end{minipage}
   \begin{minipage}{0.5\textwidth}
    \begin{center}
     \includegraphics[scale=1.2]{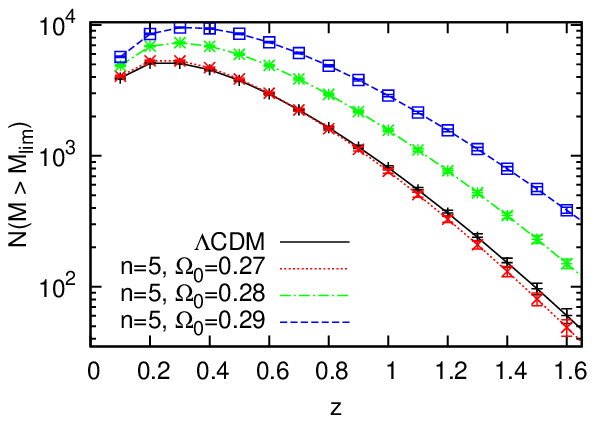}
    \end{center}
   \end{minipage}
  \end{tabular}
  \caption{The expected number of clusters $N(M>M_{\rm lim})$
  in the redshift bins of $\Delta z=0.1$, in the SPT survey for 
  the $\mathrm{\Lambda CDM}$ model with $\sigma_8=0.8$ and the kinetic 
  braiding model $n=5$. 
  The left panel compares the $\Lambda$CDM model with 
  $\sigma_8=0.8$ and the kinetic braiding model with 
  $\sigma_8=0.75,~0.8,$ and $0.85$, respectively,
  with fixing  $\Omega_0h^2=0.1334$ for all cases.
  The right panel compares the $\Lambda$CDM model with 
  the kinetic braiding model with $\Omega_0=0.27,~0.28$ 
  and $0.29$, respectively, with fixing the initial
  amplitude of the fluctuation so as to be $\sigma_8=0.8$ 
  in the $\mathrm{\Lambda CDM}$ model.
  \label{fig11}}
\end{figure}

We consider the expected redshift distributions of clusters 
in the upcoming survey, the Southern Polar Telescope (SPT) 
Sunyaev Zeldovich (SZ) survey with a limiting flux density of 
$f_{\nu_0, lim }=5 \mathrm{mJy}$ at $\nu_0=150 \mathrm{GHz}$ and a sky coverage of 
${\cal A}=4 \times 10^3 \mathrm{deg^2}$. 
The relation between the limiting flux and halo mass in the SPT survey 
\cite{Fedeli} is given by
\begin{eqnarray}
  f_{\nu_0, lim }=\frac{2.592 \times 10^8 \mathrm{mJy}}{d_A^2(z)}
\left(\frac{M_{\rm lim}}{10^{15}M_{\odot}}\right)^{1.876} 
\left(\frac{H(z)}{H_0}\right)^{2/3},
\label{LimitingFlux}
\end{eqnarray}
where $d_A(z)$ is the angular diameter distance. % $d_A(z)=d_L(z)/(1+z)^2$. 
In figure~\ref{fig11}, we show the expected redshift distributions of 
clusters in the SPT survey in redshift bins of $\Delta z=0.1$ 
including Poisson errors for clusters above the limiting halo mass $M_{\rm lim}$, 
which is determined by eq.~(\ref{LimitingFlux}). 
In the left panel, we compare the $\mathrm{\Lambda CDM}$ model
and the kinetic braiding model with $n=5$, where
we fixed $\Omega_0h^2=0.1334$, and adopted
$\sigma_8=0.8$ for the $\mathrm{\Lambda CDM}$ 
model and $\sigma_8=0.75,~0.8,$ and $0.85$, respectively, 
for the kinetic braiding model. 
In the right panel, we compare the $\mathrm{\Lambda CDM}$ 
model and the kinetic braiding model with $n=5$ and 
$\Omega_0=0.27,~0.28$, and $0.29$, respectively, in which
we fixed the initial amplitude of the power 
spectrum for the kinetic braiding model so that
$\sigma_8=0.8$ for the $\mathrm{\Lambda CDM}$ model. 

As demonstrated in figure~\ref{fig11}, the redshift distribution of 
the high-redshift clusters is practically useful to distinguish 
between the $\Lambda$CDM model and the kinetic braiding model,
depending on the cosmological parameters and the model parameters.
However, figure~\ref{fig11} also demonstrates that the
degeneracy between $\Omega_0$, $n$ and $\sigma_8$ is problematic. 
Therefore, in order to distinguish between the $\Lambda$CDM model 
and the kinetic braiding model,
we need an independent constraint on the amplitude of the 
perturbations on the kinetic braiding model. 
This will be
obtained by combining a measurement of the cosmic microwave
background anisotropies or a measurement of the cosmic
shear statistics from weak lensing surveys. The Fisher 
matrix analysis in section \ref{FMACC} estimates a constraint on $\Omega_0$ 
and $n$, assuming the amplitude of the cosmological perturbation
can be determined from those independent observations.

%%%%%%%%%%%%%%%%%%%%%%%%%%%%%%%%%%%%%%%%%%%%%%%%%%%%%%%%%
\subsection{Fisher Matrix Analysis}
\label{FMACC}
%%%%%%%%%%%%%%%%%%%%%%%%%%%%%%%%%%%%%%%%%%%%%%%%%%%%%%%%%
In this subsection, we investigate a constraint from the SPT survey
predicted with the Fisher matrix analysis. 
The Fisher matrix for the cluster counts (e.g., \cite{WK})
can be written as
\begin{eqnarray}
F_{ij}=\sum_m{\partial N_m\over \partial \theta_i}
{\partial N_m\over \partial \theta_j}{1\over N_m},
\end{eqnarray}
where $N_m$ is the number of galaxy clusters in the $m$th redshift bin
of the range $z_m\leq z<z_{m+1}$, 
\begin{eqnarray}
N_m=\int_{z_m}^{z_{m+1}} dz {dN \over dz}.
\end{eqnarray}
% with
% \begin{eqnarray}
% {dN(z)\over dz}={dV\over dz}\int_{M_1}^{M_2}{dn(M,z)\over dM} dM.
% \end{eqnarray}
In the Fisher matrix analysis for the cluster counts,
we consider 2 parameters, $n$ and $\Omega_0$, and 
adopt the target parameter $n=5$ and $\Omega_0h^2=0.1344$.
The other cosmological parameters are fixed as those in section 
\ref{FMAPK}.

% Fisher matrix of the 6 parameters, 
% $n$, $\Omega_0$, $b_0$, $b_1$, $b_2$ and $\sigma_v$ over
% the parameters other than  $n$ and $\Omega_0$.
% In figure~\ref{fig_fisher_ps} we assumed a redshift survey like 
% WFMOS/SuMIRe, 
% whose sky coverage is $2000$ $\mathrm{deg}^2$, 
% the minimum and the maximum redshifts are  
% $z_{\rm min}=0.8$ and $z_{\rm max}=1.6$, respectively, 
% and the averaged number density of galaxy  
% $\bar{n}_g=3 \times 10^{-4}h^3\mathrm{Mpc}^{-3}$. 
% The target model is the kinetic braiding model with 
% the target parameter $n=5$, $\Omega_0h^2=0.1344$, $h=0.7$, 
% $b_0=2.0$, $b_1=0.1$, $b_2=0.1$, and $\sigma_v=350 {\rm km/s}$.
% We assumed the the power spectrum is measured adopting the fiducial
% distance-redshift relation $s=s(z)$ of the spatially flat 
% $\Lambda$CDM model with $\Omega_0$=0.28, and $\sigma_8=0.8$.

\begin{figure}[t]
  \begin{center}
    \includegraphics[scale=1.3]{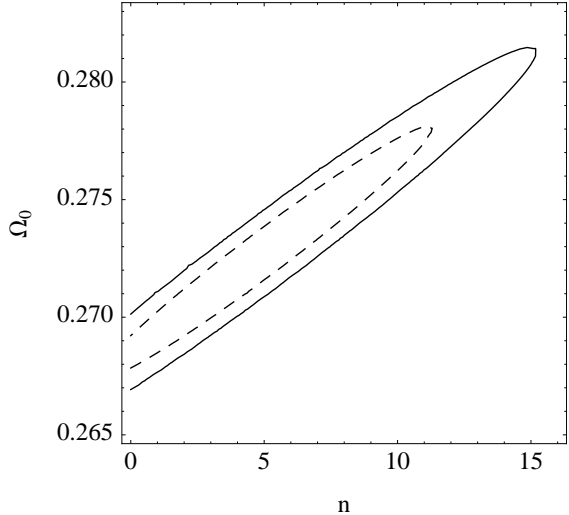}% 
  \end{center}
  \caption{
The 1-sigma (dashed curve) and 2-sigma (solid curve) contours 
in the $\Omega_0 - n$ plane from the Fisher matrix for the 
cluster distribution assuming like the 
SPT survey with redshift measurements (see section \ref{GCC}). 
The target parameter is $n=5$ and $\Omega_0h^2=0.1344$.
We fixed the spectral index $n_s=0.96$ and  
the amplitude of the initial cosmological perturbation 
so that $\sigma_8=0.8$ in the limit of the $\Lambda$CDM.
\label{fig_fisher_cc} {}}
\end{figure}

Figure~\ref{fig_fisher_cc} shows 
the 1-sigma (dashed curve) and 2-sigma (solid curve) confidence 
contours in the $n$ and $\Omega_0$ plane, 
assuming like the SPT survey 
whose sky coverage is ${\cal A}=4 \times 10^3 \mathrm{deg^2}$
and the limiting flux, as explained in the previous section.
{}From figure~\ref{fig_fisher_cc} one can see the similar feature
as that in figure~\ref{fig_fisher_ps}.  
The constraint on $n$ is not very tight.
However, the 1-sigma (2-sigma) error in determining $n$ 
is $\Delta n\sim 10$ ($15$) when $n=5$. 
Thus, it is possible to distinguish between the
kinetic braiding model and the $\Lambda$CDM model if $n\simlt 10$.
However, the constraint on the model parameter $n$ become very 
weak when the target value of $n$ becomes large because 
the kinetic braiding model approaches the $\Lambda$CDM model.

%%%%%%%%%%%%%%%%%%%%%%%%%%%%%%%%%%
\section{Conclusion}
%%%%%%%%%%%%%%%%%%%%%%%%%%%%%%%%%%
In this paper, we studied the cosmological consequences of the
kinetic braiding model, whose self-interaction term is in 
the form, $G(\phi,X)\square\phi \propto X^n\square\phi$. 
This model allows the existence of a self-accelerating solution 
without ghost and instability for perturbations.
Due to the Vainshtein mechanism, general relativity recovers on 
small scales, which ensures the consistency with the solar system 
tests. 
In this model with assuming the attractor condition 
(\ref{Att}), 
we found that the background evolution is identical
to that of the Dvali-Turner's model, which approaches the 
$\Lambda \mathrm{CDM}$ for $n$ equal to infinity. 
We also found that the linear growth history of the cosmological 
perturbation is identical to that of the $\Lambda \mathrm{CDM}$ model
for $n=\infty$.
Using the Ia supernovae data \cite{SCP} and the WMAP CMB data 
\cite{Komatsu}, we obtained a 
constraint on the model parameters of the kinetic braiding model.
The kinetic braiding model with $n=1$ requires rather high 
matter density parameter, which is consistent with the result 
of \cite{OCG}. We found the model with the higher $n$
can be consistent with the observations.
We also found that the sound velocity of the galileon field 
perturbation becomes zero for $n=\infty$. 
This prevents the perturbation of the galileon field from
propagating, which breaks the quasi-static approximation 
when $n$ is large.

By solving the full equations of the linear cosmological perturbations
numerically, we confirmed that the quasi-static approximation can 
be used for sub-horizon modes $k\simgt 0.01~h $ ${\rm Mpc}^{-1}$ 
as long as $n\simlt 10$. 
We investigated the evolution of the linear density 
perturbations as well as the spherical collapse in the nonlinear 
regime of the density perturbations, which are important 
in order to distinguish between the kinetic braiding model 
and the $\Lambda$CDM model when $n$ is small. 
We also found a useful analytic expression for the growth index, 
in a form expanded in terms of the matter density parameter.
The theoretical prediction for the large scale structure is 
confronted with the multipole power spectrum of the luminous 
red galaxy sample of the Sloan Digital Sky survey. 
The current constraint is not very tight, but a future redshift 
survey like the WFMOS/SuMIRe PFS survey would be potentially useful to 
distinguish between the kinetic braiding model 
and the $\Lambda$CDM model if $n$ is small.

Based on the quasi-static approximation and the sub-horizon approximation
for model $n\simlt10$,
we also investigated the non-linear evolution of density perturbations 
using the spherical collapse, where the nonlinear effect of the 
galileon field perturbation plays an important role through the 
Vainshtein mechanism.
We showed that the effective gravitational constant in the 
perturbation including the nonlinear effect approaches  
the usual Newton's constant after the density becomes high. 
We also investigated a prospect for the number count of galaxy clusters
in the SPT survey, which will be a useful test by combining other 
measurements like the cosmic microwave background anisotropies 
or the cosmic shear statistics of weak lensing surveys.

The kinetic braiding model in the present paper reduces to the 
$\Lambda$CDM model for $n=\infty$, which we demonstrated  
at least as for the background expansion and the linear cosmological 
perturbations. In this meaning, the kinetic gravity braiding
can be unbraided by taking a large value of $n$.
This is an interesting feature that might be useful to understand 
the origin of the cosmological constant.

%Therefore, other obsevations, such as Integrated Sachs-Wolfe effect, 
%might be a chance to constrain these model parameters.
%We shoud note that the Vainstein mechanism might no longer work
%for large $n$ and this prediction for the cluster number counts
%is only allowed for the small $n$.

\vspace{3mm}
\acknowledgments
This work was supported by Japan Society for Promotion
of Science (JSPS) Grants-in-Aid
for Scientific Research (Nos.~21540270,~21244033).
This work is also supported by JSPS 
Core-to-Core Program ``International Research 
Network for Dark Energy''.
We thank K.~Koyama, T.~Yano, G.~Nakamura, T.~Narikawa, K.~Ichiki, N. Sugiyama, 
A.~Taruya, F.~Schmidt, T.~Kobayashi, and J.~Soda for useful comments and 
discussions. We are grateful to I. Sawicki and his 
collaborators for useful comments on the investigation of the constraint
from the CMB data.
We also thank T.~Sato and G.~Huetsi for providing us with the
data of the multipole power spectrum of the SDSS LRG 
sample.
We also thank M.~Sasaki, A.~De~Felice, 
S.~Mukohyama and T.~Takahashi for useful discussions.

\appendix

%%%%%%%%%%%%%%%%%%%%%%%%%%%%%%%%%%
\section{Linear Perturbation Equations}
\label{appA}
%%%%%%%%%%%%%%%%%%%%%%%%%%%%%%%%%%
In this appendix, we summarize perturbation equations for the general 
Lagrangian (\ref{Lagrangian}).
The $(0,0)$ component of the Einstein equations is given by
\begin{eqnarray}
&&M_{\mathrm{Pl}}^2 \left[-6H(\dot{\Phi}-H\Psi)+\frac{2}{a^2}\nabla^2\Phi \right]
\nonumber
\\
&&~~
=K_{\phi}\delta\phi-K_{X}\delta X - K_{\phi X}\dot{\phi}^2\delta\phi
-K_{XX}\dot{\phi}^2 \delta X 
\nonumber
\\
&&~~
+2G_{\phi}\delta X+G_{\phi\phi}\dot{\phi}^2\delta\phi
+G_{\phi X}\dot{\phi}^2 (\delta X-3H\dot{\phi} \delta\phi)
\nonumber
\\
&&~~
-G_{X}\left[3\dot{\phi}^3\dot{\Phi}-12H\dot{\phi}^3\Psi
+9H\dot{\phi}^2\dot{\delta\phi}
-\frac{\dot{\phi}^2}{a^2}\nabla^2\delta\phi \right]
\nonumber
\\
&&~~
-3G_{XX}H\dot{\phi}^3\delta X-\delta\rho,
\label{fullPertE00apdx}
\end{eqnarray}
where $\delta X$ is defined by eq.(\ref{defdeltaX}) and $\delta \rho$
is the matter density perturbation.
The $(0,i)$ component of the Einstein equation is
\begin{eqnarray}
M_{\mathrm{Pl}}^2 \left[2(\dot{\Phi}-H\Psi)\right]
&=&-\left(K_{X}-2G_{\phi}\right)\dot{\phi}\delta\phi
\nonumber\\
&&~~
-G_X 
\dot{\phi}^2 \left(\dot{\phi}\Psi-\dot{\delta\phi}+3H\delta\phi \right)+\delta q,
\label{fullPertE0iapdx}
\end{eqnarray}
where $\delta q$ describes the velocity field of the matter component
(See below for details). 
The traceless part of the $(i,j)$ component of the Einstein equation
gives (\ref{fullPertEtl}), while
%\begin{eqnarray}
%  \Psi+\Phi = 0.
%\label{fullPertEtlapdx}
%\end{eqnarray}
the trace part yields %of the $(i,j)$ component of the Einstein equation
\begin{eqnarray}
&&M_{\mathrm{Pl}}^2 \left[2(3H^2+2\dot{H})\Psi+2H\dot{\Psi}-2\ddot{\Phi}-6H\dot{\Phi}
+\frac{1}{a^2}\nabla^2(\Psi+\Phi)  \right]
\nonumber
\\
&&~~
=K_{\phi}\delta\phi+K_{X}\delta X-2G_{\phi}\delta X-G_{\phi\phi}\dot{\phi}^2\delta\phi
-G_{\phi X}\dot{\phi}^2(\delta X+\ddot{\phi}\delta\phi)
\nonumber
\\
&&~~
+G_{X}\left[\dot{\phi}^3\dot{\Psi}+4\dot{\phi}^2\ddot{\phi}\Psi
-\dot{\phi}^2\ddot{\delta\phi}-2\dot{\phi}\ddot{\phi}\dot{\delta\phi} \right]
%\nonumber
%\\
%&&~~
-G_{XX}\dot{\phi}^2\ddot{\phi}\delta X +\delta p,
\label{fullPertEijapdx}
\end{eqnarray}
where $\delta p$ is the perturbation of pressure.
The perturbation equation for the galileon field is given by
\begin{eqnarray}
~\hspace{-1cm}
&&(2G_{\phi}-K_{X})\left[3\dot{\phi}\dot{\Phi}-\dot{\phi}\dot{\Psi}-2(\ddot{\phi}
+3H\dot{\phi})\Psi+\ddot{\delta\phi}+3H\dot{\delta\phi}
-\frac{1}{a^2}\nabla^2\delta\phi \right]
\nonumber
\\
&&~~~~
+(K_{\phi\phi}-K_{\phi\phi X}\dot{\phi}^2+G_{\phi\phi\phi}\dot{\phi}^2)\delta\phi
-(K_{XXX}+3G_{XXX}H\dot{\phi})\dot{\phi}^2\ddot{\phi}\delta X
\nonumber
\\
&&~~~~
-(K_{\phi X}-2G_{\phi\phi})\left[\delta X+(\ddot{\phi}
+3H\dot{\phi})\delta\phi \right]
-K_{\phi XX}\dot{\phi}^2(\delta X+\ddot{\phi}\delta\phi)
\nonumber
\\
&&~~~~
+K_{XX}\left[\dot{\phi}^3\dot{\Psi}+(5\ddot{\phi}+3H\dot{\phi})\dot{\phi}^2\Psi
-\dot{\phi}^2\ddot{\delta\phi}-3(\ddot{\phi}+H\dot{\phi})\dot{\phi}\dot{\delta\phi} \right]
\nonumber
\\
&&~~~~
-G_{X}\biggl[3\dot{\phi}^2\ddot{\Phi}+6(\ddot{\phi}+3H\dot{\phi})\dot{\phi}\dot{\Phi}
-9H\dot{\phi}^2\dot{\Psi}
\nonumber
\\
&&~~~~
-12\biggl\{(\dot{H}+3H^2)\dot{\phi}^2
+2H\dot{\phi}\ddot{\phi} \biggr\}\Psi
-\frac{\dot{\phi}^2}{a^2}\nabla^2\Psi
+6H\dot{\phi}\ddot{\delta\phi}
\nonumber
\\
&&~~~~
+6\biggl\{H\ddot{\phi}+(\dot{H}+3H^2)\dot{\phi} \biggr\}\dot{\delta\phi}
-\frac{2}{a^2}(\ddot{\phi}+2H\dot{\phi})\nabla^2\delta\phi
 \biggr]
\nonumber
\\
&&~~~~
-G_{\phi X}\biggl[\dot{\phi}^3(3\dot{\Phi}+\dot{\Psi})+6(\ddot{\phi}
-H\dot{\phi})\dot{\phi}^2\Psi-\dot{\phi}^2\ddot{\delta\phi}
-(4\ddot{\phi}-3H\dot{\phi})\dot{\phi}\dot{\delta\phi}
\nonumber
\\
&&~~~~
+3\biggl\{2H\dot{\phi}\ddot{\phi}+(\dot{H}+3H^2)\dot{\phi}^2 \biggr\}\delta\phi
-\frac{\dot{\phi}^2}{a^2}\nabla^2\delta\phi\biggr]
\nonumber
\\
&&~~~~
-G_{XX}\biggl[3\dot{\phi}^3\ddot{\phi}\dot{\Phi}-3H\dot{\phi}^4\dot{\Psi}
-3\biggl\{8H\dot{\phi}^3\ddot{\phi}+(\dot{H}+3H^2)\dot{\phi}^4\biggr\}\Psi
\nonumber
\\
&&~~~~~~~~~~~~~~~~
+3H\dot{\phi}^3
\ddot{\delta\phi}
+3\biggl\{5H\dot{\phi}^2\ddot{\phi}+(\dot{H}+3H^2)\dot{\phi}^3\biggr\}
\dot{\delta\phi}-\frac{\dot{\phi}^2\ddot{\phi}}{a^2}\nabla^2\delta\phi\biggr]
\nonumber
\\
&&~~~~~
+G_{\phi\phi X}\dot{\phi}^2\biggl[\delta X+(\ddot{\phi}-3H\dot{\phi})
\delta\phi\biggr]
%\nonumber
%\\
%&&~~~~~
+G_{\phi XX}\dot{\phi}^2 \left[(\ddot{\phi}-3H\dot{\phi})\delta X
-3H\dot{\phi}\ddot{\phi}\delta\phi \right]=0.
\label{fullPertFieldapdx}
\end{eqnarray}

Let us summarize the perturbations of the matter component, whose
evolution equation is given by the 
energy-momentum conservation $\nabla_\mu T^{\mu\nu}{}=0$. 
The perturbation of the energy-momentum tensor is given by 
$\delta T^0_0=-\delta\rho$, $\delta T^0_i=\partial_i \delta q$, and 
$\delta T^i_j=\delta p \delta^i_j$, respectively.
The energy-momentum conservation yields
\begin{eqnarray}
  &&\dot{\delta\rho}+3H(\delta \rho+\delta p) +3{\dot \Phi}(\rho+p)+\frac{\nabla^2}{a^2}\delta q=0, \\
  &&\dot{\delta q}+3H\delta q +(\rho+p)\Psi+\delta p=0. 
\end{eqnarray}
For the matter perturbations, we can set $\delta p=0$. 
It is convenient to define the quantity $\delta q = -\rho v$.
In terms of the density contrast $\delta \equiv \delta \rho /\rho$ and $v$, 
these equations can be written as
\begin{eqnarray}
  &&\dot{\delta}+3{\dot \Phi}-\frac{\nabla^2}{a^2}v=0, \\
\label{EMC1}
  &&\dot{v}=\Psi.
\label{EMC2}
\end{eqnarray}
Taking the time derivative of (\ref{EMC1}) and using (\ref{EMC2}), 
we obtain the evolution equation for the matter density perturbation,
\begin{eqnarray}
  \ddot{\delta}+2H\dot{\delta}=\frac{\nabla^2}{a^2}\Psi-3\ddot{\Phi}
-6H\dot{\Phi}.
\end{eqnarray}
Finally, we introduce the gauge-invariant density perturbation, 
$\Delta_c \equiv \delta+3Hv$, and the new variable, $B \equiv -\Phi+Hv$. 
Then, the matter perturbation obeys the following equation,
\begin{eqnarray}
  \ddot{\Delta}_c+2H\dot{\Delta}_c=\frac{\nabla^2}{a^2}\Psi+3\ddot{B}+6H\dot{B}.
\label{fullDensityEvolution}
\end{eqnarray}


\begin{thebibliography}{99}
%************************OBSERVATIONS******************************%
\bibitem{Riess}
  A. G. Riess et al., {\it Observational Evidence from Supernovae for an Accelerating Universe and a Cosmological Constant, Astron. J.} {\bf 116} (1998) 1009 
\bibitem{Perlmutter}
  S. Perlmutter et al., {\it Measurements of Omega and Lambda from 42 High-Redshift Supernovae, Astropys. J.} {\bf 517} (1999) 565 
\bibitem{Spergel}
  D. N. Spergel et al., {\it Three-Year Wilkinson Microwave Anisotropy Probe (WMAP) Observations: Implications for Cosmology, Astropys. J. Suppl.} 
{\bf 170} (2007) 377 
\bibitem{Komatsu}
  E. Komatsu, et~al.,{\it Seven-year Wilkinson Microwave Anisotropy Probe (WMAP) Observations: Cosmological Interpretation, Astrophys. J. Suppl.}
{\bf 192} (2011) 18
\bibitem{Percival}
  B. Reid, et al., {\it Cosmological constraints from the clustering of the Sloan Digital Sky Survey DR7 luminous red galaxies, Mon. Not. Roy. Astron. Soc.} {\bf 404} (2010) 60 
\bibitem{sloan}
  M. Tegmark et al., {\it Cosmological constraints from the SDSS luminous red galaxies, Phys. Rev.} {\bf D 74} (2006) 123507 
\bibitem{Allen}
 K. Vanderlinde, et al., {\it Galaxy Clusters Selected with the Sunyaev-Zel'dovich Effect from 2008 South Pole Telescope Observations, Astrophys. J.}  {\bf 722} (2010) 1180 
\bibitem{Rapetti}
  D. Rapetti, S. W. Allen and J. Weller, {\it Constraining dark energy with X-ray galaxy clusters, supernovae and the cosmic microwave background,  Mon.~Not.~Roy.~Astron.~Soc.} {\bf 360} (2005) 555 
\bibitem{PeeblesRatra}
  P. J. E. Peebles and B. Ratra, {\it The cosmological constant and dark energy, Rev. Mod. Phys.} {\bf 75} (2003) 559 
\bibitem{Padmanabhan}
  T. Padmanabhan, {\it Cosmological constant-the weight of the vacuum, Phys. Rept.} {\bf 380} (2003) 235 
\bibitem{Weinberg}
  S. Weinberg, {\it The cosmological constant problem, Rev. Mod. Phys.} 
{\bf 61} (1989) 1 
\bibitem{Weinberg2}
  S. Weinberg, {\it The Cosmological Constant Problems}, astro-ph/0005265
%**********************MODIFIED GRAVITY******************************%
\bibitem{Amendola}
L. Amendola, {\it Scaling solutions in general nonminimal coupling theories, Phys. Rev.} {\bf D 60} (1999) 043501 
\bibitem{Uzan}
J. P. Uzan, {\it Cosmological scaling solutions of nonminimally coupled scalar fields, Phys. Rev. }  {\bf D 59} (1999) 123510 
\bibitem{Chiba}
T. Chiba, {\it Quintessence, the gravitational constant, and gravity, Phys. Rev.}  {\bf D 60} (1999) 083508 
\bibitem{Bartolo}
N. Bartolo and M. Pietroni, {\it Scalar-tensor gravity and quintessence, Phys. Rev. } {\bf D 61} (2000) 023518 
\bibitem{Perrotta}
F. Perrotta, C. Baccigalupi and S. Matarrese, {\it Extended quintessence,
Phys. Rev.} {\bf D 61} (2000) 023507 
\bibitem{Carroll}
 S. M. Carroll, V. Duvvuri, M. Trodden and M. S. Turner, {\it Is cosmic speed-up due to new gravitational physics?, Phys. Rev. } 
{\bf D 70} (2004) 043528 
\bibitem{Nojiri}
 S. Nojiri and S. D. Odintsov, {\it Modified gravity with negative and positive powers of curvature: Unification of inflation and cosmic acceleration, Phys. Rev.} {\bf D 68} (2003) 123512 
\bibitem{Capozziello}
 S. Capozziello, S. Carloni and A. Troisi, {\it Quintessence without scalar fields, Recent Res. Dev. Astron. Astrophys.} {\bf 1} (2003) 625
\bibitem{HuSawicki}
 W. Hu and I. Sawicki, {\it Models of f(R) cosmic acceleration that evade solar system tests, Phys. Rev.} {\bf D 76} (2007) 064004 
\bibitem{Starobinsky}
 A. A. Starobinsky, {\it Disappearing cosmological constant in f(R) gravity, JETP Lett.} {\bf 86} (2007) 157 
\bibitem{Tsujikawafr}
 S. Tsujikawa, {\it Observational signatures of f(R) dark energy models that satisfy cosmological and local gravity constraints, Phys. Rev.} 
{\bf D 77} (2008) 023507 
\bibitem{Nojirib}
 S. Nojiri and S. Odintsov, {\it Unifying inflation with ĩCDM epoch in modified f(R) gravity consistent with Solar System tests, Phys. Lett.}  
{\bf B 657} (2007) 238 
\bibitem{DGP1}
 G. R. Dvali, G. Gabadadze and M. Porrati, {\it Metastable gravitons and infinite volume extra dimensions, Phys. Lett.} 
{\bf B 484} (2000) 112 
\bibitem{DGP2}
 G. R. Dvali, G. Gabadadze and M. Porrati, {\it 4D gravity on a brane in 5D Minkowski space, Phys. Lett.} {\bf B 485} (2000) 208 
\bibitem{Song}
Y-S. Song, I. Sawicki and W. Hu, {\it Large-scale tests of the Dvali-Gabadadze-Porrati model, Phys. Rev.} {\bf D 75} (2007) 064003 
\bibitem{Maartens}
 R. Maartens and E. Majerotto, {\it Observational constraints on self-accelerating cosmology, Phys. Rev.} {\bf D 74} (2006) 023004 
\bibitem{KM}
 R. Maartens and K. Koyama, {\it Brane-World Gravity,  Living Rev. Rel.} {\bf 13} (2010) 5 
\bibitem{Schmidt}
 F. Schmidt, A. Vikhlinin and W. Hu, {\it Cluster constraints on f(R) gravity, 
 Phys. Rev.}  {\bf D 80} (2009) 083505 
\bibitem{frr}
 K. Yamamoto, G. Nakamura, G. Hutsi, T. Narikawa and T. Sato, {\it Constraint on the cosmological f(R) model from the multipole power spectrum of the SDSS luminous red galaxy sample and prospects for a future redshift survey,
 Phys. Rev.} {\bf D 81} (2010) 103517 
\bibitem{GC}
  N. Chow and J. Khoury, {\it Galileon cosmology, Phys. Rev.} 
{\bf D 80} (2009) 024037
\bibitem{SAUGC}
  F. P. Silva and K. Koyama, {\it Self-accelerating universe in Galileon cosmology, Phys. Rev.} {\bf D 80} (2009) 121301 
\bibitem{ELCP}
  T. Kobayashi, H. Tashiro and D. Suzuki, {\it Evolution of linear cosmological perturbations and its observational implications in Galileon-type modified gravity, Phys. Rev.}  {\bf D 81} (2010) 063513 
\bibitem{CEGH}
  T. Kobayashi, {\it Cosmic expansion and growth histories in Galileon scalar-tensor models of dark energy, Phys. Rev.}  {\bf D 81} (2010) 103533 
\bibitem{GGC}
  A. De Felice and S. Tsujikawa, {\it Generalized Galileon cosmology}, arXiv:1008.4236 
\bibitem{GBDT}
  A. De Felice and S. Tsujikawa, {\it Generalized Brans-Dicke theories, 
JCAP} {\bf 07} (2010) 024 
\bibitem{DPGMGT}
  A. De Felice, S. Mukohyama and S. Tsujikawa, {\it Density perturbations in general modified gravitational theories, Phys. Rev.}  
{\bf D 82} (2010) 023524 
\bibitem{CG}
  C. Deffayet, G. Esposito-Farese and A. Vikman, {\it Covariant Galileon, Phys. Rev.}  {\bf D 79} (2009) 084003 
\bibitem{GGIR}
  R. Gannouji and M. Sami, {\it Galileon gravity and its relevance to late time cosmic acceleration, Phys. Rev.}  {\bf D 82} (2010) 024011 
\bibitem{MGALG}
  A. Ali, R. Gannouji and M. Sami, {\it Modified gravity a la Galileon: Late time cosmic acceleration and observational constraints, Phys. Rev.} {\bf D 82} (2010) 103015 
\bibitem{CCGF}
  A. De Felice and S. Tsujikawa, {\it Cosmology of a Covariant Galileon Field, Phys. Rev. Lett.} {\bf 105} (2010) 111301 
\bibitem{OCG} 
  S. Nesseris, A. De Felice and S. Tsujikawa, {\it Observational constraints on Galileon cosmology, Phys. Rev.}  {\bf D 82} (2010) 124054 
\bibitem{CSTOG}
  D. F. Mota, M. Sandstad and T. Zlosnik, {\it Cosmology of the selfaccelerating third order Galileon}, arXiv:1009.6151    
\bibitem{FeliceTsujikawa}
A. De Felice, R. Kase, S. Tsujikawa, 
{\it Matter perturbations in Galileon cosmology}, arXiv:1011.6132
\bibitem{Deffayet} 
  C. Deffayet, O. Pujolas, I. Sawicki and A. Vikman, {\it Imperfect dark energy from kinetic gravity braiding, JCAP} {\bf 10} (2010) 026 
\bibitem{DDEF}
  C. Deffayet, S. Deser and G. Esposito-Farese, {\it Generalized Galileons: All scalar models whose curved background extensions maintain second-order field equations and stress tensors, Phys. Rev.}  {\bf D 80} (2009) 064015 
\bibitem{GALMG}
  A. Nicolis, R. Rattazzi and E. Trincherini, {\it Galileon as a local modification of gravity, Phys. Rev.}  {\bf D 79} (2009) 064036
\bibitem{RFFGM}
  C. Burrage and D. Seery, {\it Revisiting fifth forces in the Galileon model, JCAP} {\bf 08} (2010) 011 
\bibitem{SSSDBI}
  G. L. Goon, K. Hinterbichler and M. Trodden, {\it Stability and superluminality of spherical DBI galileon solutions}, arXiv:1008.4580
\bibitem{DBIGR}
  C. de Rham and A. J. Tolley, {\it DBI and the Galileon reunited, JCAP} {\bf 05} (2010) 015 
\bibitem{BGT1}
  A. Padilla, P. M. Saffin and S. Zhou, {\it Bi-galileon theory I: motivation and formulation, JHEP} {\bf 12} (2010) 031
\bibitem{BGT2}
  A. Padilla, P. M. Saffin and S. Zhou, {\it Bi-galileon theory II: phenomenology}, arXiv:1008.3312
\bibitem{BTJC}
  E. Dyer and K. Hinterbichler, {\it Boundary terms and junction conditions for the DGP Pi-Lagrangian and galileon, JHEP} {\bf 11} (2009) 059 
\bibitem{APS}
  A. Padilla, P. M. Saffin and S. Zhou, {\it Multi-galileons, solitons and Derrick's theorem}, arXiv:1008.0745
\bibitem{KMD}
  K. Hinterbichler, M. Trodden and D. Wesley, {\it Multifield Galileons and higher codimension branes, Phys. Rev.} {\bf D 82} (2010) 124018
\bibitem{MKJMT}
  M. Andrews, K. Hinterbichler, J. Khoury and M. Trodden, {\it Instabilities of Spherical Solutions with Multiple Galileons and SO(N) Symmetry}, arXiv:1008.4128
\bibitem{CSG}
  C. Deffayet, S. Deser and G. Esposito-Farese, {\it Arbitrary p-form 
Galileons, Phys. Rev.}  {\bf D 82} (2010) 061501 
\bibitem{GA}
  E. Babichev, {\it Galileon accretion}, arXiv:1009.2921
\bibitem{GI}
  C. Burrage, C. de Rham, D. Seery and A. J. Tolley, {\it Galileon inflation, JCAP} {\bf 01} (2011) 014
\bibitem{GIKobayashi}
  T. Kobayashi, M. Yamaguchi and J. Yokoyama, {\it 
Inflation Driven by the Galileon Field, Phys. Rev. Lett.} 
{\bf 105} (2010) 231302 
\bibitem{GIKobayashiII}
  K. Kamada, T. Kobayashi, M. Yamaguchi, J. Yokoyama, 
{\it Higgs G-inflation}, arXiv:1012.4238
\bibitem{MK}
  S. Mizuno and K. Koyama,{\it Primordial non-Gaussianity from the DBI Galileons, Phys. Rev.} {\bf 82} (2010) 103518
\bibitem{DGP3}
A. Nicolis and R. Rattazzi, {\it Classical and Quantum Consistency of the DGP Model, JHEP} {\bf 06} (2004) 059
\bibitem{DGP4}
M. A. Luty, M. Porrati and R. Rattazzi, {\it Strong interactions and stability in the DGP model, JHEP} {\bf 09} (2003) 029
\bibitem{Vainshtein}
  A. I. Vainshtein, {\it To the problem of nonvanishing gravitation mass, Phys. Lett.}  {\bf B 39} (1972) 393
%**********************************************%
\bibitem{DT2003}
  G. Dvali and M. S. Turner, {\it Dark energy as a modification of the Friedmann equation}, astro-ph/0301510
\bibitem{KoyamaDGPn}
 K. Koyama, {\it Structure formation in modified gravity models, JCAP} 
{\bf 03} (2006) 017
%**********************************************%
\bibitem{Uzanb}
  J-P. Uzan, {\it The acceleration of the universe and the physics behind it}, 
astro-ph/0605313
\bibitem{YS2007}
  K. Yamamoto, D. Parkinson, T. Hamana, R.C. Nichol and Y. Suto, {\it Optimizing future imaging survey of galaxies to confront dark energy and modified gravity models, Phys. Rev.} {\bf D 76} (2007) 023504 [{\it Erratum ibid} {\bf D 76} (2007) 129901]
%%%%%%%%%%%%%%%%%%%%%%%%%%%%%%%%%%%%%%%%%%%%%%%%%%%%%%%%%%%%%%%
\bibitem{KoyamaMaartens}
 K. Koyama and R. Maartens, {\it Structure formation in the Dvali Gabadadze Porrati cosmological model, JCAP} {\bf 01} (2006) 016 
%**********************SUPERNOVAE******************************%
\bibitem{SCP}
  R. Amanullah et al., {\it Spectra and Hubble Space Telescope Light Curves of Six Type Ia Supernovae at 0.511$<$z$<$1.12 and the Union2 Compilation, Astrophys. J.} {\bf 716} (2010) 712
\bibitem{WangMukherjee}
Y. Wang and P. Mukherjee, {\it Observational constraints on dark energy and cosmic curvature, Phys. Rev.} {\bf D 76} (2007) 103533
\bibitem{HuSugiyama}
W. Hu and N. Sugiyama, {\it Small-Scale Cosmological Perturbations: an Analytic Approach, Astrophys. J.} {\bf 471} (1996) 542 
%**********************GROWTH INDEX******************************%
\bibitem{Creminelli09112701}
P. Creminelli, G. D'Amico, J. Norena, L. Senatore, F. Vernizzi, 
{\it Spherical collapse in quintessence models with zero speed of sound, 
JCAP} 03 (2010) 027
%**********************GROWTH INDEX******************************%
\bibitem{Peebles}
  P.~J.~E. Peebles, {\it Large-Scale Structure of the Universe}, 
Princeton University Press (1980)
\bibitem{Linder}
  E.~V. Linder, {\it Cosmic growth history and expansion history, Phys. Rev.} 
{\bf D 72} (2005) 043529 
%**********************POWER SPECTRUM******************************%
\bibitem{Lindergr}
E. V. Linder, {\it Redshift distortions as a probe of gravity, Astropart. Phys.} {\bf 29} (2008) 336 
\bibitem{Guzzo}
 L. Guzzo et~al., {\it A test of the nature of cosmic acceleration using galaxy redshift distortions, Nature} {\bf 451} (2008) 541
\bibitem{Yamamoto08}
 K. Yamamoto, T. Sato and G. H\"{u}tsi, {\it Testing General Relativity with the Multipole Spectra of the SDSS Luminous Red Galaxies, Prog. Theor. Phys.}~{\bf 120} (2008) 609 
\bibitem{WSP}
 M. White, Y. Song and W. J. Percival, {\it Forecasting cosmological constraints from redshift surveys, Mon. Not. Roy. Astron. Soc.} {\bf 397} (2009) 1348
\bibitem{Reyes}
 R. Reyes, et al., {\it Confirmation of general relativity on large scales from weak lensing and galaxy velocities, Nature} {\bf 464} (2010) 256 
\bibitem{Hirano}
K. Hirano, {\it Observational tests of Galileon gravity with growth rate}, arXiv:1012.5451
\bibitem{devonv}
 T. Sato, G. H\"{u}etsi and K. Yamamoto, {\it Deconvolution of Window Effect in Galaxy Power Spectrum Analysis, Prog. Theor. Phys.} {\bf 125} (2011) 187
%**********************POWER SPECTRUM******************************%
\bibitem{Jennings}
  E. Jennings, C. M. Baugh and S. Pascoli, {\it Modelling redshift space distortions in hierarchical cosmologies, Mon.~Not.~Roy.~Astron.~Soc.}
{\bf 410} (2010) 2081
\bibitem{PD} J. A. Peacock and S. J. Dodds,  {\it Non-linear evolution of cosmological power spectra, Mon.~Not.~Roy.~Astron.~Soc.} {\bf 280} (1996) L1
\bibitem{EHu}
  D. J. Eisenstein and W. Hu, {\it Baryonic Features in the Matter Transfer Function, Astrophys. J.} {\bf 496} (1998) 605 
\bibitem{Scoccimarro} 
  R. Scoccimarro, {\it Redshift-space distortions, pairwise velocities, and nonlinearities, Phys. Rev.} {\bf D 70} (2004) 083007 
\bibitem{BPH}
 W. E. Ballinger, J. A. Peacock and A. F. Heavens, 
 {\it Measuring the cosmological constant with redshift surveys, Mon. Not. Roy. Astron. Soc.}  {\bf 282} (1996) 877 
\bibitem{Koyama09} K. Koyama, A. Taruya and T. Hiramatsu, {\it Nonlinear evolution of the matter power spectrum in modified theories of gravity, 
Phys.~Rev.} {\bf D 79} (2009) 123512
% DGP quasi-nonlinear 
% \bibitem{Oyaizu08a} H. Oyaizu, Phys.~Rev.~D {\bf 78},123523 (2008) 
% \bibitem{Oyaizu08b} H. Oyaizu, M. Lima, W. Hu, Phys.~Rev.~D {\bf 78}, 
%123524 (2008)
\bibitem{Schmidt09a}F. Schmidt, {\it Cosmological simulations of normal-branch braneworld gravity, Phys.~Rev.} {\bf D 80} (2009) 123003
   % DGP N-body simulation
\bibitem{ScoccimarroNbody}
 R. Scoccimarro, {\it tLarge-scale structure in brane-induced gravity. I. Perturbation theory, Phys. Rev.}  {\bf D 80} (2009) 104006
\bibitem{sumire}
 H. Aihara, talk at the IPMU international conference on 
dark energy: lighting up the darkness!, Kashiwa, Japan, 2009
\bibitem{WFMOS} B. Bassett, R. C. Nichol and D. J. Eisenstein, {\it WFMOS: Sounding the dark cosmos, Astronomy and Geophysics} {\bf 46} (2005) 5.26
\bibitem{Nakamichi} K. Yamamoto, M. Nakamichi, A. Kamino, B. A. Bassett
and H. Nishioka, {\it  Measurement of the Quadrupole Power Spectrum in the Clustering of the 2dF QSO Survey, Publ. Astron. Soc. Japan} {\bf 58}  (2006) 93 

%************************Non-linear evolution******************************%
\bibitem{Koyama07}
  K. Koyama and F. P. Silva, {\it Nonlinear interactions in a cosmological background in the Dvali-Gabadadze-Porrati braneworld, Phys. Rev.} {\bf D 75} (2007) 084040
\bibitem{Lue04}
  A. Lue, R. Scoccimarro and G. D. Starkman, {\it Probing Newton's constant on vast scales: Dvali-Gabadadze-Porrati gravity, cosmic acceleration, and large scale structure, Phys. Rev.}  {\bf D 69} (2004) 124015
\bibitem{Schmidt10}
  F. Schmidt, W. Hu and M. Lima, {\it Spherical collapse and the halo model in braneworld gravity, Phys. Rev.}  {\bf D 81} (2010) 063005 
%************************Spherical collapse******************************%
\bibitem{Gunn}
  J. E. Gunn and J. R. Gott, {\it On the Infall of Matter Into Clusters of Galaxies and Some Effects on Their Evolution, Astrophys. J.} {\bf 176} (1972) 1
\bibitem{Lahav91}
O. Lahav, P. B. Lilje, J. R. Primack and M. J. Rees, {\it Dynamical effects of the cosmological constant, Mon. Not. Roy. Astron. Soc.} {\bf 251} (1991) 128
\bibitem{Wang98}
L. Wang and P. J. Steinhardt, {\it Cluster Abundance Constraints for Cosmological Models with a Time-varying, Spatially Inhomogeneous Energy Component with Negative Pressure, Astrophys. J.} {\bf 508} (1998) 483
\bibitem{Horellou05}
C. Horellou and J. Berge, {\it Dark energy and the evolution of spherical overdensities,  Mon. Not. Roy. Astron. Soc.} {\bf 360} (2005) 1393 
\bibitem{Basilakos09}
S. Basilakos, J. C. B. Sanchez and L. Perivolaropoulos, {\it Spherical collapse model and cluster formation beyond the ĩ cosmology: Indications for a clustered dark energy?, Phys. Rev.}  {\bf D 80} (2009) 043530
\bibitem{mota06}
M. Manera and D. F. Mota, {\it Cluster number counts dependence on dark energy inhomogeneities and coupling to dark matter, Mon. Not. Roy. Astron. Soc.} {\bf 371}, (2006) 1373
%\bibitem{mota08}
%D. F. Mota, {\it Probing Dark Energy at Galactic and Cluster Scales, JCAP} {\bf 0809} (2008) 006
%************************Energy Nonconservation******************************%
\bibitem{Maor05}
I. Maor and O. Lahav, {\it On virialization with dark energy, JCAP} 
{\bf 07} (2005) 003 
%************************MASS FUNCTION******************************%
\bibitem{PressSchechter}
  W. H. Press and P. Schechter, {\it Formation of Galaxies and Clusters of Galaxies by Self-Similar Gravitational Condensation,   Astrophys. J.} 
{\bf 187} (1974) 425
\bibitem{Fedeli}
  C. Fedeli, L. Moscardini, and S. Matarrese, {\it The clustering of galaxy clusters in cosmological models with non-Gaussian initial conditions: predictions for future surveys, Mon. Not. Roy. Astron. Soc.} {\bf 397} (2009) 1125 
%************************Fisher******************************%
\bibitem{yamamoto2003}
K. Yamamoto, {\it Optimal Weighting Scheme in Redshift-Space Power Spectrum Analysis and a Prospect for Measuring the Cosmic Equation of State, Astrophys. J.} {\bf 595} (2003) 580
\bibitem{WK}
S. Wang, J. Khoury, Z.Haiman, {\it Constraining the evolution of dark energy with a combination of galaxy cluster observables, Phys. Rev.} {\bf D 70} (2004) 123008 
%%%%%%%%%%%%%%%%%%%%%%%%%%%%%%%%%%%%%%%%%%%%%%%%%%%%%%%%%%%%%%%%%%%%%%%%%%%
\end{thebibliography}
\end{document}